\documentclass[graybox]{svmult}

\usepackage{mathptmx}       
\usepackage{helvet}         
\usepackage{courier}        
\usepackage{type1cm}        
\usepackage{makeidx}         
\usepackage{graphicx}        
\usepackage{multicol}        
\usepackage[bottom]{footmisc}

\makeindex             

\begin{document}

\title*{Topology of quantum vacuum}
\author{G.E. Volovik}
\institute{
 Low Temperature Laboratory, Aalto University, P.O. Box 15100, FI-00076 AALTO, Finland \\
L.D. Landau Institute for Theoretical Physics, Kosygina 2, 119334 Moscow, Russia 
\\
\email{volovik@boojum.hut.fi}
}
%
%
\maketitle

\abstract*{ Topology in momentum space is the main characteristics of  the ground states of a system at zero temperature, the quantum vacua. The gaplessness of fermions in bulk, on the surface or inside the vortex core is protected by topology, and is not sensitive to details of the microscopic physics (atomic or trans-Planckian). Irrespective of the deformation of the parameters of the microscopic theory, the energy spectrum of these fermions remains strictly gapless. This solves the main hierarchy problem in particle physics: for fermionic vacua with Fermi points the masses of elementary particles are naturally small. The quantum vacuum of Standard Model is one of the representatives of topological matter alongside with topological superfluids and superconductors, topological insulators and semi-metals, etc. There is a number of of topological invariants in momentum space of different dimensions. They determine universality classes of the topological matter and the type of the effective theory which  emerges at low energy. In many cases they also give rise to emergent symmetries, including the effective Lorentz invariance, and  emergent phenomena such as effective gauge and gravitational fields.  The topological invariants in extended momentum and coordinate space determine the bulk-surface and bulk-vortex correspondence. They connect the momentum space topology in bulk with the real space. These invariants determine the gapless fermions living on the surface of a system or in the core of topological defects (vortices, strings, domain walls, solitons, monopoles, etc.).  The momentum space topology gives some lessons for quantum gravity. In effective gravity emerging at low energy, the collective variables are the tetrad field and spin connections, while the metric is the composite object of tetrad field. This suggests that the Einstein-Cartan-Sciama-Kibble theory with torsion field is more relevant. There are also several scenarios of Lorentz invariance violation governed by topology, including splitting of Fermi point and development of the Dirac points with quadratic and cubic spectrum. The latter leads to the natural emergence of the Ho\v{r}ava-Lifshitz gravity.
}

\abstract{Topology in momentum space is the main characteristics of  the ground states of a system at zero temperature, the quantum vacua. The gaplessness of fermions in bulk, on the surface or inside the vortex core is protected by topology, and is not sensitive to details of the microscopic physics (atomic or trans-Planckian). Irrespective of the deformation of the parameters of the microscopic theory, the energy spectrum of these fermions remains strictly gapless. This solves the main hierarchy problem in particle physics: for fermionic vacua with Fermi points the masses of elementary particles are naturally small. The quantum vacuum of Standard Model is one of the representatives of topological matter alongside with topological superfluids and superconductors, topological insulators and semi-metals, etc. There is a number of of topological invariants in momentum space of different dimensions. They determine universality classes of the topological matter and the type of the effective theory which  emerges at low energy. In many cases they also give rise to emergent symmetries, including the effective Lorentz invariance, and  emergent phenomena such as effective gauge and gravitational fields.  The topological invariants in extended momentum and coordinate space determine the bulk-surface and bulk-vortex correspondence. They connect the momentum space topology in bulk with the real space. These invariants determine the gapless fermions living on the surface of a system or in the core of topological defects (vortices, strings, domain walls, solitons, monopoles, etc.).  The momentum space topology gives some lessons for quantum gravity. In effective gravity emerging at low energy, the collective variables are the tetrad field and spin connections, while the metric is the composite object of tetrad field. This suggests that the Einstein-Cartan-Sciama-Kibble theory with torsion field is more relevant. There are also several scenarios of Lorentz invariance violation governed by topology, including splitting of Fermi point and development of the Dirac points with quadratic and cubic spectrum. The latter leads to the natural emergence of the Ho\v{r}ava-Lifshitz gravity.
}

\section{Introduction}\label{sec:introduction}

Topological approach seeks to recover  the Standard Model of particle physics and general relativity as the low-energy phenomena  emerging in the complicated microscopic media called the quantum vacuum. Momentum space topology distributes the quantum vacua into universality classes, which are characterized by different types of topological invariants, expressed in terms of propagators -- Green's functions. The similar invariants characterize the ground states of topological materials:  topological superfluids, superconductors, insulators, semi-metals, etc. The only difference between 
the topological materials and the quantum vacuum of  the Standard Model is that the latter obeys several symmetries, which are absent in condensed matter physics. These are
the Lorentz invariance and different types of gauge symmetry. However, in principle, the symmetries of the Standard Model can arise as the emergent phenomena.  This is precisely what happens in the 
 universality class of vacua, which are characterized by the existence of the topologically protected Fermi point in momentum space -- the point where the energy of the fermionic excitations is necessarily nullified due to non-zero topological invariant. Such point exists in the vacuum of the Standard Model in the symmetric phase above the electroweak interaction, and also in some topological materials, whose ground states also belong to the Fermi point universality class. These are superfluid $^3$He in the phase A, and topological semimetals. That is why we call the symmetric phase of the Standard Model
 as quantum vacuum in the semi-metal state.
 At low energy, i.e. in the vicinity of the point node in the spectrum of fermionic excitations, these excitations behave as relativistic particles -- chiral (left handed or right handed) Weyl fermions -- interacting with effective gauge fields and gravitational fields.
In principle, all the ingredients of Standard Model and general relativity can be recovered as phenomena naturally emerging in the background of the vacuum (ground state), which belongs to the Fermi point universality class. 

 Here we consider also the other universality classes of topological media, illustrating them on condensed matter and particle physics examples. Among them is the vacuum of Standard Model
 below the electroweak transition, where the momentum-space topology prescribes the appearance of masses for all fermionic excitations: quarks and leptons. This massive phase of the Standard Model vacuum is also topolgically nontrivial: it belongs to universality class, which includes such topological media as superfluid $^3$He in the phase B and three-dimensional topological insulators obeying time reversal symmetry. The fermionic spectrum in all these systems is fully gapped, but there are topologically protected gapless (massless) states on the surface of these media or at the interfaces. That is why we call the massive phase of the Standard Model
 as the quantum vacuum in the insulating state. 
 
 Another interesting class of topological media includes the vacua where the fermionic excitations are exotic gapless Dirac particles with non-relativistic quadratic, cubic or higher order spectrum. The effective quantum electrodynamics, which emerges in such vacua, experiences the 
so-called anisotropic scaling, in which the space and time transform in different ways. The topological  mechanism of emergent anisotropic scaling can serve as a source of the recently suggested 
quantum gravity at short distances (the so-called Ho\v{r}ava-Lifshitz gravity).

Some sections are devoted to the fermionic bound states leaving on the boundary of the topological media, on the interfaces between the vacua with different topological invariants, or inside the topological defects, such as strings in relativistic theories and vortices in superfluids and superconductors.  The topology can say may or may not these bound states have exactly zero energy. In superfluids and supercondictors such zero-energy states behave as Majorana fermions, which are still elusive in particle physics. The topological consideration of bound states suggests another scenario of emergence of the chiral fermions of the Standard Model: they may emerge as the fermion zero modes at the interface between two different insulating quantum vacua in the 4+1 space-time.

\section{Quantum vacuum as topological medium}\label{sec:introduction}

\subsection{Symmetry vs topology}\label{sec:Symmetry_topology}

 There is a fundamental interplay of symmetry and topology in physics, both in condensed matter and relativistic quantum fields. Traditionally the main role was played by symmetry: gauge symmetry of Standard Model and GUT; symmetry classification of condensed matter systems such as solid and liquid crystals, magnets, superconductors and superfluids; universality classes of spontaneously broken symmetry phase transitions; etc.   The last decades demonstrated the opposite tendency in which topology is becoming primary being the main characteristics of  quantum vacua in relativistic
 quantum fields and their condensed-matter counterparts -- ground states of the condensed-matter systems at   $T=0$, see reviews  
 \cite{Volovik2003,Volovik2007,HasanKane2010,Xiao-LiangQi2011}
 and earlier papers
\cite{NielsenNinomiya1981,VolovikMineev1982,NielsenNinomiya1983,Avron1983,Semenoff1984,NiuThoulessWu1985,So1985,IshikawaMatsuyama1986,IshikawaMatsuyama1987,SalomaaVolovik1988,Haldane1988,Volovik1988,Yakovenko1989,Horava2005}.

Topology describes the properties of a system, which are insensitive to the details of the microscopic physics (atomic physics in condensed matter and Planck-scale physics of quantum vacuum). 
The momentum-space topology, which we discuss in this Chapter, determines universality classes of the topological media  and the effective quantum field theory which  emerges in such media at low energy and low temperature, including the type of the energy spectrum of
fermionic excitations (in condensed matter systems fermionic excitations play the role of elementary particles). The topology also gives rise to emergent symmetry, the symmetry which did not exist
on a microscopic level. Examples are provided by the point nodes (zeroes) in the energy spectrum of fermionic excitations: some of the nodes   are protected by topology,
i.e. they are robust to the deformations of the system. Close to such nodes the spectrum
of fermions becomes ``relativistic'', i.e. the spectrum forms the Dirac cone, and fermionic excitations behave as Weyl, Dirac or Majorana particles.  As a result the effective relativistic quantum field theory  emerges, in which bosonic collective modes give rise to effective gauge field and effective metric. All this
is the consequence of the topological theorem -- the Atiyah-Bott-Shapiro construction  \cite{Horava2005}. 

Among the existing and potential representatives of topological materials one 
can find those in which the spectrum of fermionic excitations has a  gap (the gap corresponds to 
mass of a relativistic particle). These systems include 3D topological band insulators  \cite{HasanKane2010}; fully gapped superluid $^3$He-B \cite{SalomaaVolovik1988,Schnyder2008,Kitaev2009};  2D materials exhibiting intrinsic (i.e. without external magnetic field) quantum Hall and spin-Hall effects, such as gapped graphene \cite{Haldane1988}; thin film of superluid $^3$He-A and quasi 2D planar phase of triplet superfluid
\cite{Volovik1988,VolovikYakovenko1989,Yakovenko1989}; and chiral superconductor 
Sr$_2$RuO$_4$
\cite{Mackenzie2003}. These materials have the topological properties similar to that of the quantum vacuum of Standard Model in its massive phase, i.e. in the state below the electroweak transition, where
all elementary particles are massive \cite{Volovik2010a}. The latter state can be called the ``insulating'' state  of
the relativistic quantum vacuum.

The gapless topological media are represented by  superluid $^3$He-A \cite{Volovik2003}; 
topological semimetals \cite{Abrikosov1971,Abrikosov1998,Burkov2011,XiangangWan2011}; gapless graphene 
\cite{Ryu2002,Manes2007,Volovik2007,Vozmediano2010,Cortijo2011}; nodal cuprate  superconductors
\cite{Ryu2002}; and noncentrosymmetric   
\cite{SchnyderRyu2010,SchnyderBrydonTimm2011} superconductors.  These materials are similar to the quantum vacuum of Standard Model in the state above   the electroweak transition, where all elementary particles are massless
\cite{Volovik2003,Volovik2010a}.  This state can be called the ``semimetal'' phase  of
the relativistic quantum vacuum.

\begin{figure*}
\includegraphics[width=\textwidth]{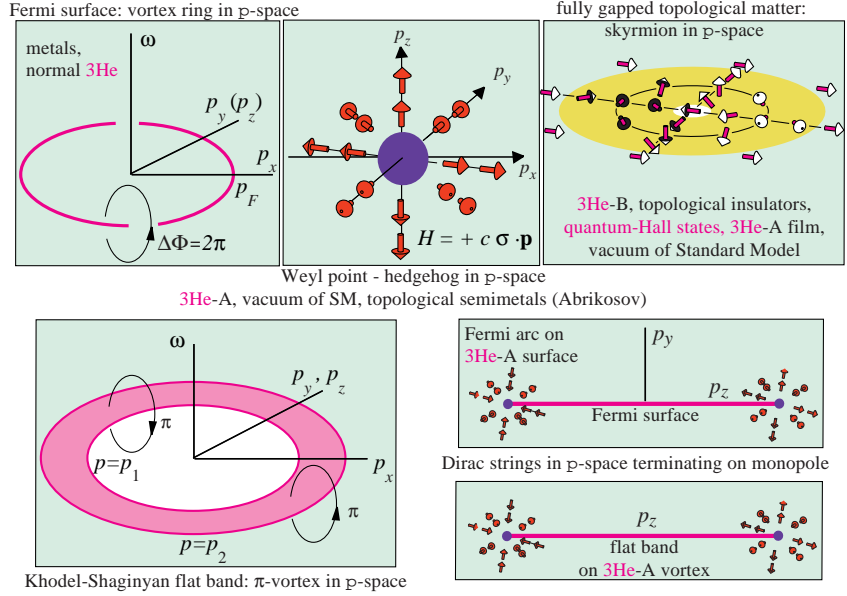}
\caption{Topological matter, represented in terms of topological objects in momentum space.
({\it top left}): Fermi surface is the momentum-space analog of the vortex line: the phase of the Green's function changes by $2\pi$ around the element of the line in $(\omega, {\mathbf p})$-space. ({\it top middle}):  Fermi point (Weyl point) is the counterpart of a hedgehog and a magnetic monopole. The hedgehog in this figure has integer topological charge $N=+1$, and close to this Fermi point the fermionic quasiparticles behave as Weyl fermions. Nontrivial topological charges in terms of Green's functions support the stability of the Fermi surfaces and Weyl points with respect to perturbations including interactions \cite{Volovik2003,EssinGurarie2011}. In terms of the Berry phase \cite{Berry1984}
the Fermi point
represents the ${\bf p}$-space counterpart of Dirac magnetic monopole with unobservable Dirac string
(see Ref. \cite{Volovik1987}  and Fig. 11.4 in \cite{Volovik2003}). ({\it top right}):  Topological insulators and fully gapped topological superfluids/superconductors are textures in momentum space: they have no singularities in the Green's function and thus no nodes in the energy spectrum in the bulk. This figure shows a skyrmion in the two-dimensional momentum space, which characterizes two-dimensional topological insulators exhibiting intrinsic quantum Hall or spin-Hall effect. ({\it bottom left}):  Flat band emerging in strongly interacting systems  \cite{Khodel1990}. This dispersionless Fermi band is analogous to a soliton terminated by half-quantum vortices: the phase of the Green's function changes by $\pi$ around the edge of the flat  band   \cite{NewClass}.  
({\it bottom right}): Fermi arc on the surface of $^3$He-A \cite{Tsutsumi2011} and of topological semi-metals with Weyl points \cite{XiangangWan2011,Burkov2011} and flat band inside the vortex core of $^3$He-A \cite{Volovik2010b}  serve as the momentum-space analog of a Dirac string terminating on a monopole. The Fermi surface formed by the surface bound states terminates on the points where the spectrum of zero energy states merge with the continuous spectrum in the bulk, i.e. with the Weyl points.}
\label{TopUnClasses}
\end{figure*}

\subsection{Green's function vs order parameter}\label{sec:Green_function}

Topology operates in particular with integer numbers -- topological invariants or topological charges -- which do not change under small deformation of the system. The topological invariants appear in very different situations.
The commonly known are topological charges, which describe the topological defects -- the {\it inhomogeneous} objects such as quantized vortices, dislocations, domain walls, hedgehogs, solitons, etc.  in condensed matter systems  or the cosmic strings, magnetic monopoles, instantons, etc. in particle physics. The topological charge protects the topological defects from destruction, while the summation law of the topological charges regulates processes of scattering, merging, splitting, and other transformations of these defects in their dynamics.

Here we consider the topological invariants of different type: they describe the {\it homogeneous} ground states or
the {\it homogeneous} quantum vacua. The conservation of these topological charges in particular protects 
 from destruction different types of nodes in the fermionic energy spectrum of a system.  
 The node in the spectrum is an object in momentum space, and thus we call these invariants
 as momentum-space invariants, as distinct from the real-space invariants describing 
 topological defects. 
In other words, the real-space invariants describes the topologically nontrivial inhomogeneous configurations of the order parameter fields $\Psi({\bf r},t)$ in space-time, while the momentum-space invariants describe the nontrivial momentum-space configuration of the propagator -- Green's function $G({\bf p},\omega)$ or other response function, which characterizes 
the homogeneous ground state of a system (the vacuum state) 
\cite{Volovik2003,Horava2005,EssinGurarie2011,Zubkov2012b}. 
The Green's function is generally a matrix with spin
indices. In addition, it may have the band indices (in the case of 
electrons in the periodic potential of crystals).

 Due to topological stability, the nodes in the spectrum survive when the interaction between the fermions is introduced and/or modified. Thus the momentum-space topology is the main reason why there are gapless   quasiparticles in condensed matter  and (nearly) massless  elementary particles in our Universe. Topology explains the masslessness of the quantum vacuum in its semimetal phase, and allows to solve the hierarchy problem in particle physics: if the masslessness (or gaplessness) of the quantum vacuum is not protected, the natural values of the masses of elementary particles are on the order of the Planck energy scale, which is in huge disagreement with masses of elementary particles in the insulating vacuum. However, if masses appear as a result of the spontaneously broken symmetry
in the initially gapless vacuum, they can be small.

 The momentum-space topological invariants are in many respects similar to their real-space counterparts, which describe topological defects in condensed matter systems and quantum vacuum
(see Fig. \ref{TopUnClasses}).
In particular,  the 
point node in the spectrum (the Fermi point or Weyl point, see Sec.  \ref{sec:Vacuum_semi-metal}) is the counterpart of  the real-space point defects, such as hedgehog  in ferromagnets or magnetic monopole in particle physics (Fig. \ref {TopUnClasses} {\it top center}). The summation law of the momentum-space topological charges regulates processes of scattering, merging, splitting, and other transformations of the nodes in spectrum
during deformation of the vacuum state. Another example is the Fermi surface in metals, which we consider in Sec. \ref{FS}. It is topologically stable, because it is analogous to the vortex loop in superfluids or superconductors (Fig. \ref{TopUnClasses} {\it top left}). 

The fully gapped topological matter, such as topological insulators and fully gapped topological superfluids have no nodes in the spectrum and thus no singularities in the Green's function. They are the counterpart of the real-space objects which do not have singularities, but still are topologically non-trivial -- the so-called textures or skyrmions.
Thus the topological insulator and the vacuum of Standard Model in insulating phase can be considered as the momentum-space skyrmion  (Fig.~1 {\it top right}).

The further extension of topology to the combined real + momentum space allows us
to consider  topologically protected  spectrum of  the so called fermion zero modes -- fermions living on the real-space topological objects such as domain walls, strings and monopoles.
The fermion zero modes are protected  by topological invariants expressed in terms of the Green' s function in the extended phase space, $G(\omega,{\bf p};t,{\bf r})$
\cite{GrinevichVolovik1988,SalomaaVolovik1988,Volovik1989c,Volovik2003,SilaevVolovik2010,EssinGurarie2011}.

\subsection{Fermi surface as topological object}
\label{FS}

\begin{figure}[t]
\centerline{\includegraphics[width=0.5\linewidth]{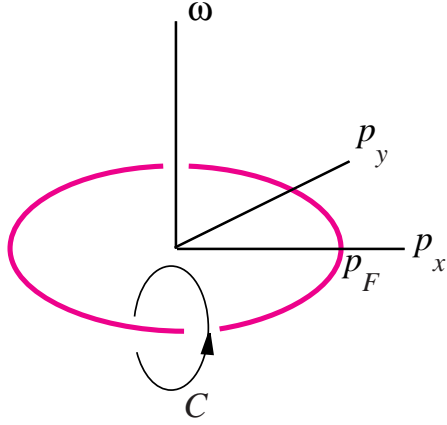}}
\medskip
\caption{Nodes of co-dimension 1 in 2+1 systems. Green's function has singularities on line $\omega=0$, $p_x^2+p_y^2=p_F^2$ in the three-dimensional space $(\omega,p_x,p_y)$.
 Stability of Fermi surface is protected by the invariant (\ref{InvariantForFS}) which is represented by  integral  over an arbitrary contour $C$ around the Green's function singularity.
 This is applicable to nodes of co-dimension 1 in any $D$+1 dimension. For  $D=3$ the nodes form conventional Fermi surface in metals and in normal $^3$He.}
\label{FermiSurfaceQPTFig}
\end{figure}

We shall start with gapless vacua -- the vacua whose fermionic excitations has no gap.  In the gapless media the fermionic degrees of freedom are not frozen out at low temperature. This is very important for practical purposes in electronic devices and for the existence of life in our Universe. 
It is the topology which provides the protection for the gaplessness in condensed matter systems and masslessness of elementary particles in Standard Model.

For the topological classification of the gapless vacua, the Green's function is considered on imaginary frequency axis. 
On the real axis singularities are always present, for example in the simplest case of non-interacting fermions the poles in the Green's function, $G=1/\left(\omega- \epsilon({\bf p}\right))$, are present even in the fully gapped system. To distinguish between the gapped and gapless states of the vacuum, and also between different types of gapless vacua, we should move to the imaginary axis.
Then the simplest Green's function becomes $G=1/\left(i\omega- \epsilon({\bf p}\right))$, which may have singularity at  $\omega=0$ and $\epsilon({\bf p})=0$. The latter is possible only for the gapless vacua, where the energy of excitations $\epsilon({\bf p})$ has zeroes. 
Thus zeroes in the spectrum of fermionic excitations are manifested as singularities in the Green's function, and we are interested in such singularities in the Green's function, which are protected by topology, i.e. are described by topological charge and thus are robust to perturbations of the vacuum state.
The transition to imaginary axis thus allows us to consider only the relevant singularities in the Green's function and to avoid the trivial singularities on the mass shell. 

Zeroes in the energy spectrum of fermionic excitations may form different $p$-dimensional manifolds
in momentum space: there can be zero-dimensional points $p=0$, one-dimensional lines of zeroes $p=1$,  
two-dimensional surfaces $p=2$ and even the whole three-dimensional bands
with zero energy $p=3$. Since the dimension $D$ of the momentum space can be also different, for 
topological classification it is more instructive to use the co-dimension of zeroes.
 By co-dimension we denote $D-p$, i.e. the dimension $D$ of ${\bf p}$-space minus dimension $p$ of the nodes in energy spectrum. 

We start with zeroes of co-dimension 1. This refers to two-dimensional Fermi surface in three-dimensional metal ($3-2=1$), D=1 Fermi line in D=2 systems  ($2-1=1$) and Fermi point in $D=1$ systems ($1-0=1$).
 The general
analysis \cite{Horava2005} demonstrates that topologically stable  
nodes of co-dimension 1 are described by the group $Z$ of integers. The  corresponding winding number
$N$ is expressed analytically  in terms of the Green's function
\cite{Volovik2003}:
\begin{equation}
N={\bf tr}~\oint_C {dl\over 2\pi i}  G(\omega,{\bf p})\partial_l
G^{-1}(\omega,{\bf p})~.
\label{InvariantForFS}
\end{equation}
Here the integral is taken over an arbitrary contour $C$ around the Green's function singularity
in  the $D+1$
momentum-frequency space. See Fig. \ref{FermiSurfaceQPTFig} for $D=2$. Example of the Green's function in any dimension $D$ is scalar function
$G^{-1}(\omega,{\bf p})=i\omega - v_F(|{\bf p}|-p_F)$. For $D=2$, the singularity with winding number $N=1$ is on the line $\omega=0$, $p_x^2+p_y^2=p_F^2$, which represents the one-dimensional Fermi surface.

Due to nontrivial topological invariant, Fermi surface survives the perturbative interaction and exists even in systems without poles in the Green's function and quasiparticles are not well defined. The systems without poles include the so-called marginal and Luttinger liquids in condensed matter and the so-called unparticles in
relativistic theories (see Sec. \ref{sec:Emergent_matter}).

 \section{Vacuum in a semi-metal state}
 \label{sec:Vacuum_semi-metal}

Here we consider the topology of the quantum vacuum of Standard Model in its symmetric phase above the electroweak transition, where all fermions are massless Weyl fermions. Standard Model vacuum obeys the relativistic invariance and thus cannot belong to the universality class of vacua with Fermi surface:  the latter violates the Lorentz symmetry. 
The gapless (massless) phase of Standard Model can be consistent with  Lorentz symmetry only if the vacuum of Standard Model belongs to  the universality class
of vacua, which is characterized by point nodes in the fermionic spectrum -- the Fermi points.
The Fermi point has the co-dimension 3. The Fermi point can be of the chiral type, when the  fermionic excitations near the Fermi point behave as relativistic left-handed or right-handed Weyl fermions, such as left neutrino, left electron, left quarks and their right-handed counterparts 
\cite{Froggatt1991,Volovik2003}. In this case the Fermi point is called the Weyl point.  The Fermi point can be also of the Dirac type, when the  fermionic excitations behave as massless Dirac fermions.
There is also the class of quantum vacua in which the nodal point obeys the $Z_2$ topology, with the topological charge obeying the summation rule $1+1=0$. In such vacuum the fermionic excitations behave as relativistic massless Majorana particles  \cite{Horava2005}.

\subsection{Fermi points in 3+1 vacua}
\label{sec:Fermi_points}

Fermi point is the Green's function singularity described by the following topological invariant expressed via integer valued integral over the surface $\sigma$ around the singular point   in the 4-momentum space   $p_\mu=(\omega,{\bf p})$ \cite{Volovik2003}:
 \begin{equation}
N = \frac{e_{\alpha\beta\mu\nu}}{24\pi^2}~
{\bf tr}\int_\sigma   dS^\alpha
~ G\partial_{p_\beta} G^{-1}
G\partial_{p_\mu} G^{-1} G\partial_{p_\nu}  G^{-1}\,.
\label{MasslessTopInvariant3D}
\end{equation}
If the invaraint (\ref{MasslessTopInvariant3D}) is nonzero, the Green's function has a singularity inside the surface $\sigma$, and this means that fermions are gapless. The typical singularities
have topological charge $N =+1$ or $N =-1$.
 In the vicinity  of the Fermi point with such topological invariant 
 the spectrum of fermionic excitations reproduces the spectrum the chiral Weyl fermions, right-handed or left-handed respectively. 
 This is the consequence of the  so-called Atiyah-Bott-Shapiro construction \cite{Horava2005},  which leads to the following general form of expansion of the inverse fermionic propagator near the Fermi point  with $N =+1$ or $N =-1$:
\begin{equation}
G^{-1}(p_\mu)=e_\alpha^\beta\Gamma^\alpha(p_\beta-p_\beta^{(0)})+ \cdots \,.
\label{Atiyah-Bott-Shapiro}
\end{equation}
Here $\Gamma^\mu=(1,\sigma_x,\sigma_y,\sigma_z)$ are Pauli matrices (or Dirac matrices in the more general case); the expansion parameters are the vector 
$p_\beta^{(0)}$ indicating the position of the Fermi point in momentum space where the Green's function has a singularity, and  the matrix $e_\alpha^\beta$; ellipsis denote higher order terms in expansion.  
The expansion results in the linear spectrum of fermionic excitations, which induces the effective Lorentz invariance.

 \subsection{Emergent relativistic fermionic matter}
 \label{sec:Emergent_matter}

By continuously adiabatic deformation one may transform equation (\ref{Atiyah-Bott-Shapiro}) to the equation which describes the relativistic Weyl fermions
\begin{equation}
G^{-1}(p_\mu)=i\omega +N{\mbox{\boldmath$\sigma$}}\cdot{\bf p} + \cdots ~~,~~N=\pm 1\,.
\label{Weyl}
\end{equation}
Here the position of the Fermi point is shifted to $p_\beta^{(0)}=0$; the matrix $e_\alpha^\beta$ is deformed to unit matrix;  and ellipsis denote higher order terms in $\omega$ and ${\bf p}$. This means that close to the Fermi point with $N=+1$, the low energy fermionic excitations behave as right handed relativistic particles, while the Fermi point with $N=-1$ gives rise to the left handed particles. 
They obey the following effective Weyl Hamiltonian
\begin{equation}
H_{\rm eff}= N{\mbox{\boldmath$\sigma$}}\cdot{\bf p}  ~~,~~N=\pm 1\,.
\label{WeylEffective}
\end{equation}
\begin{figure}
\centerline{\includegraphics[width=0.3\linewidth]{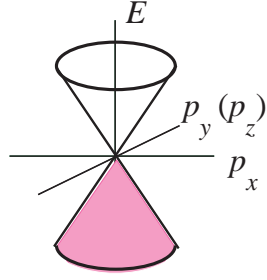}}
\caption{Weyl point as the conical point of level crossing. The negative energy branch occupied by the fermionic excitations touches the positive energy branch.
}  
\label{ConicalPoint} 
\end{figure}

The spectrum of this Hamiltonian is shown in Fig. \ref{ConicalPoint} , which demonstrates that the Fermi or Weyl point is an example of the exceptional point of level crossing  analyzed by von Neumann and Wigner \cite{Neumann1929}. This analysis demonstrates that two branches of spectrum, which have the same symmetry, may touch each other at the conical (or diabolical)  point in the three-dimensional space of parameters. In other words the
twofold degeneracy has co-dimension = 3. In our case the two branches correspond to positive and negative energy states, while the three parameters are the components of momentum  $p_x$, $p_y$ and $p_z$
\cite{Avron1983,Volovik1987}. Topological invariants for points at which the branches of spectrum merge were introduced by Novikov \cite{Novikov1981}.

However, this is not the whole story. Topological invariant of the Weyl point does not change 
if the Green's function is multiplied by any power of $p^2+\omega^2$, which
violates the simple pole structure of the propagator in Eq.(\ref{Weyl}). 
Such violation may occur due to the infrared divergences, which take place in quantum field theories. In this case  in the vicinity of Fermi point one has
\begin{equation}
G(p_\mu)\propto \frac{-i\omega +N{\mbox{\boldmath$\sigma$}}\cdot{\bf p}}{\left(p^2+\omega^2\right)^{\gamma}}  ~~,~~N=\pm 1\,,
\label{WeylUnparticle}
\end{equation}
with  $\gamma\neq 1$. In particle physics, the hypothetical particles described by the modified Green's function
with  $\gamma\neq 1$ are called unparticles  \cite{Georgi2007, LuoZhu2008}. This modification does not change the topology of the propagator: the topological charge of singularity is $N$ for arbitrary parameter  $\gamma$  \cite{Volovik2007}.
 For fermionic unparticles one has  
$\gamma=5/2 -d_U$, where $d_U$ is the scale dimension of the quantum field.

 The main property of the vacua with Weyl points is that according to (\ref{Weyl}), close to the Weyl points the massless relativistic fermions emerge. This is consistent with the fermionic content of our Universe, where all the elementary particles -- left-handed and right-handed quarks and leptons -- are Weyl fermions. Such a coincidence demonstrates that the vacuum of Standard Model is the topological medium of special  universality -- the class of vacua which have topologically
 protected  Fermi points. This solves the hierarchy problem, since the  value 
of the masses of elementary particles in the vacua of this universality class is identically zero.

Let us suppose for a moment, that there is no topological invariant which protects massless fermions,
and the vacuum of our Universe belongs to the class of the fully gappedvacua.  Then the natural masses of  fermions must be on the order of Planck energy scale:  $M \sim E_{\rm P}\sim 10^{19}$ GeV.  In such a natural  Universe, where all masses are of order $E_{\rm P}$, all fermionic degrees of freedom are completely frozen out because of the  Bolzmann factor $e^{-M/T}$, which is about $e^{-10^{16}}$   at the temperature corresponding to the  highest energy reached in accelerators. There is no  fermionic matter in such a Universe at low energy.  That we survive in our Universe is not the result of the anthropic principle (the latter chooses the Universes which are fine-tuned for life but have an extremely low probability). Our Universe is also natural and its vacuum is generic, but it belongs to the universality class of vacua with Fermi points. In such vacua the masslessness of fermions is protected by topology (combined with symmetry, see below).

 \subsection{Emergent gauge fields}
 \label{sec:Emergent_gauge}

The vacua with Fermi-point suggest a particular mechanism for emergent symmetry. The Lorentz symmetry is simply the result of the linear expansion: this symmetry becomes better and better when the Fermi point is approached and the non-relativistic higher order terms in Eq.(\ref{Weyl}) may be neglected. This expansion  demonstrates the emergence of the relativistic spin, which is described by the Pauli matrices. It also demonstrates how gauge fields and gravity emerge together with chiral fermions. 
The expansion parameters  
$p_\beta^{(0)}$ and $e_\alpha^\beta$ may depend on the space and time coordinates and they actually represent collective dynamic bosonic fields in the vacuum with Fermi point. 
The vector field  $p_\beta^{(0)}$ in the expansion plays the role of   the effective $U(1)$ gauge field $A_\beta$ acting on  fermions.

For the more complicated Fermi points with $\left|N\right|>1$ the shift $p_\beta^{(0)}$ becomes the matrix field; it gives rise to effective non-Abelian (Yang-Mills)    $SU(N)$ gauge fields emerging in the vicinity of Fermi point, i.e. at low energy \cite{Volovik2003}. For example, the Fermi point with $N=2$ may give rise to the effective $SU(2)$ gauge field in addition
to the effective $U(1)$ gauge field
\begin{equation}
G^{-1}(p_\mu)=e_\alpha^\beta\Gamma^\alpha \left(p_\beta-g_1A_\beta -g_2 {\bf A}_\beta\cdot{\mbox{\boldmath$\tau$}}\right)+~{\rm higher~order~terms}  \,,
\label{SU2}
\end{equation}
where 
${\mbox{\boldmath$\tau$}}$ are Pauli matrices corresponding to the emergent isotopic spin.

 \subsection{Emergent gravity} 
 \label{sec:Emergent_gravity}

 The matrix field $e_\alpha^\beta$ in (\ref{SU2}) acts on the (quasi)particles as  the field of vierbein, and thus describes the emergent dynamical gravity field. As a result, close to the Fermi point,  matter fields 
(all ingredients of Standard Model: chiral fermions and Abelian and non-Abelian gauge fields) 
emerge  together with geometry, relativistic spin, Dirac  matrices,  and physical laws:  Lorentz and gauge  invariance, equivalence principle, etc. This is the result of the natural coarse graining: in the low energy corner, when the Fermi point is approached, the huge  number of microscopic (Planckian) degrees of freedom of the physical quantum vacuum is reduced to rather few low-energy degrees of freedom (chiral fermions, gauge fields and
gravity field). All other degrees of freedom are highly massive and are completely frozen out at low energy and temperature.

 In such vacua, gravity emerges together with matter.   If this Fermi point mechanism of emergence of physical laws works for our Universe, then the so-called  ``quantum gravity''  does not exist. The gravitational degrees of freedom become separated from all other degrees of freedom of quantum vacuum only at low energy.  
  
In this scenario, classical gravity is a natural macroscopic phenomenon emerging in the low-energy corner of the microscopic quantum vacuum, i.e. it is a typical and actually inevitable consequence of the  coarse graining procedure discussed above. It is possible  to quantize gravitational waves to obtain their quanta -- gravitons, since in the low energy corner the results of microscopic and effective theories coincide. It is also possible to obtain some (but not all) quantum corrections to Einstein equation and to extend classical gravity to the semiclassical level.  But one  cannot  obtain ``quantum gravity'' 
(i.e. microscopic physics of the vacuum) by  quantization of Einstein equations for the gravitational degrees of freedom, since all other degrees of freedom of the quantum vacuum will be missed in this procedure.  

\subsection{Topological invariant protected by symmetry in Standard Model}
\label{TopSemi-metal}

Standard Model contains equal number $n_R =n_L$ of right and  left Weyl fermions, 8 right and 8 left in each generation: $n_R =n_L =8n_g$, where $n_g$  is the number of generations (we do not consider Standard Model with Majorana fermions, and assume that in the insulating state of Standard Model neutrinos are Dirac fermions).
For such Standard Model the total topological charge, obtained after trace is taken over all fermionic flavors in invariant (\ref{MasslessTopInvariant3D}), vanishes, $N=8n_g-8n_g=0$. Thus the invariant (\ref{MasslessTopInvariant3D}) cannot protect nodes in the spectrum of Standard Model fermions. This would mean that any interaction between the fermions may make them massive.

However, in the symmetric phase of Standard Modle above the electroweak transition this does not happen. This is because there is another topological invariant, which takes into account the symmetry of the vacuum. The gapless state of the vacuum with $N=0$ can be  protected by the following
topological invariant which is supported by symmetry \cite{Volovik2003}:
 \begin{equation}
N_K = \frac{e_{\alpha\beta\mu\nu}}{24\pi^2}~
{\bf tr}\left[K\int_\sigma   dS^\alpha
~ G\partial_{p_\beta} G^{-1}
G\partial_{p_\mu} G^{-1} G\partial_{p_\nu}  G^{-1}\right]\,.
\label{MasslessTopInvariantStandard Model}
\end{equation}
where $K_{ij}$ is the matrix of symmetry transformation, which either commutes or anticommutes
with the Green's function matrix. 
In the symmetric phase of Standard Model there are two relevant symmetries, both are the  $Z_2$ groups, $K^2=1$. One of them is the center subgroup of $SU(2)_L$ gauge group of weak rotations of left fermions, where the element $K$ is the gauge rotation by angle $2\pi$, $K=e^{i\pi {\check \tau}_{3L}}$. The other one is the group of the hypercharge rotation be angle $6\pi$, 
$K=e^{i6\pi Y}$. In the  $G(224)$ Pati-Salam extension of the $G(213)$ group of Standard Model, this symmetry comes as combination of the $Z_2$ center group of the $SU(2)_R$ gauge group for right fermions, $e^{i\pi {\check \tau}_{3R}}$, and  the element $e^{3\pi i(B-L)}$ of the
$Z_4$ center group of the $SU(4)$ color group -- the $P_M$ parity (on the importance of the discrete  groups in particle physics see 
\cite{PepeaWieseb2007,Kadastik2009} and references therein).
Each of these two $Z_2$ symmetry operations   changes sign of left spinor, but does not influence the right particles. Thus these matrices are diagonal, $K_{ij}={\rm diag}(1,1,\ldots, -1,-1,\ldots)$, with eigen values 1 for right fermions and  $-1$ for left fermions.

In the symmetric phase of Standard Model, both matrices commute with the Green's function matrix $G_{ij}$, 
as a result $N_K$ in (\ref{MasslessTopInvariantStandard Model}) is topological invariant: it is robust to deformations of Green's function which preserve the symmetry $K$.  The value of this invariant $N_K= 16 n_g$, which means that all
$16 n_g$ fermions are massless. 

Topological invariant protected by symmetry is responsible  also for the effect of chiral anomaly -- the anomalous nonconservation of a chiral current \cite{Adler1969,BellJackiw1969}. This anomaly can be the source of the production of baryonic charge $B$ in the early Universe, which leads to the excess of matter over anti-matter in the present Universe (the so-called electroweak baryogenesis, see the recent review \cite{Canetti2012}). The baryo-production can be expressed in terms of the invariant $N_K$: in case of the creation of the baryonic charge by the hypermagnetic $U(1)$ field
 the relevant symmetry operator $K$ in Eq.(\ref{MasslessTopInvariantStandard Model}) is $K=BY^2$, where $Y$ is the generator of $U(1)$ symmetry group. The creation rate of baryons in the hypermagnetic $U(1)$ field is
  \begin{equation}
\dot B=  \frac{N_K}{4\pi^2}~{\bf B}_Y\cdot {\bf E}_Y~~,~~K=BY^2 \,.
\label{Baryoproduction}
\end{equation}
In condensed matter, the chiral anomaly generated by Weyl fermions in a crystal
has been considered in Ref.  \cite{NielsenNinomiya1983}, while the analog of the baryoproduction in Eq. (\ref{Baryoproduction}) has been experimentally tested in the superfluid with Weyl points -- the $^3$He-A \cite{Bevan1997}.

\begin{figure}
\centerline{\includegraphics[width=0.7\linewidth]{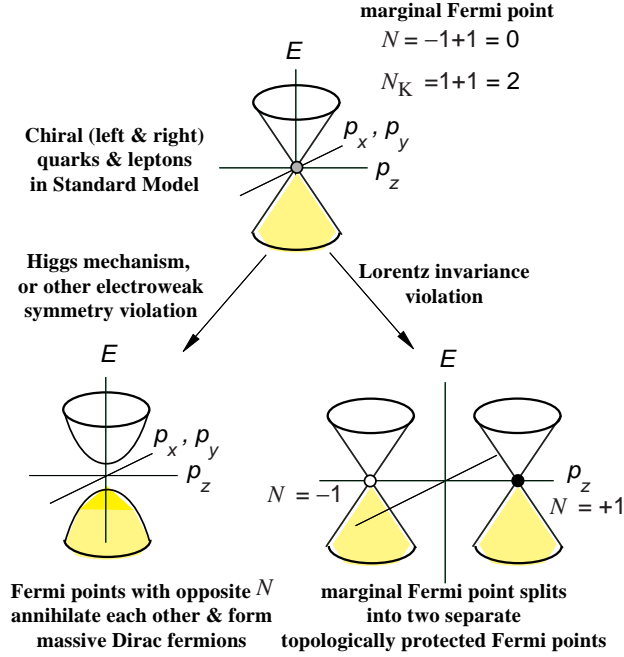}}
\caption{({\it top}) In Standard Model the Fermi points with positive $N=+1$ and negative $N=-1$ topological charges are at the same point ${\bf p}=0$, forming the marginal Fermi point with $N=0$. Symmetry $K$ between the Fermi points prevents their mutual annihilation giving rise to the topological invariant (\ref{MasslessTopInvariantStandard Model}) with $N_K=2$.  ({\it bottom left}): If symmetry $K$ is
violated or spontaneously broken,  Fermi points annihilate each other and  Dirac mass is formed. ({\it bottom right}): If Lorentz invariance is violated or spontaneously broken, the marginal Fermi point splits \cite{KlinkhamerVolovik2005a}. The topological quantum phase transition between the state with Dirac mass and the state with splitted Dirac points have been observed in cold Fermi gas
\cite{Tarruell2011}.
}  
\label{TwoScenarios} 
\end{figure}

 \subsection{Higgs mechanism vs splitting of Fermi points} 
 \label{Higgs_vs_Splitting}
 
The gapless vacuum of Standard Model is supported by combined action of topology and symmetry $K$, and also by the Lorentz  invariance which keeps all the Fermi points at ${\bf p}=0$.

Explicit violation or spontaneous breaking of one of these symmetries 
transforms the vacuum of the Standard Model
into one of the two possible vacua.  If, 
for example, the $K$  symmetry is broken, the invariant 
(\ref{MasslessTopInvariantStandard Model}) supported by this symmetry ceases to exist, and
 the Fermi point
disappears. All $16n_g$  fermions become massive  
(Fig.~\ref{TwoScenarios} {\it bottom left}). This is assumed to happen below the
symmetry breaking electroweak transition caused by Higgs mechanism
where quarks and charged leptons acquire the Dirac masses.  

If, on the other hand, the Lorentz symmetry is violated, the
marginal Fermi point splits into topologically stable Fermi
points with non-zero invariant $N$, which protects massless chiral fermions
(Fig.~\ref{TwoScenarios} {\it bottom right}). Since the invariant $N$ does not depend on symmetry, the further symmetry breaking cannot destroy the nodes.
One can speculate that in the  Standard Model the latter may happen
with the electrically neutral leptons, the neutrinos
\cite{KlinkhamerVolovik2005a}. Most interestingly, Fermi-point splitting of neutrinos
may provide a new source of T and CP violation
in the leptonic sector, which may be relevant
for the creation of the observed cosmic
matter-antimatter asymmetry \cite{Klinkhamer2006}.

  \subsection{Splitting of Fermi points and problem of generations} 
 \label{FPS_generations}

\begin{figure}[top]
\centerline{\includegraphics[width=0.8\linewidth]{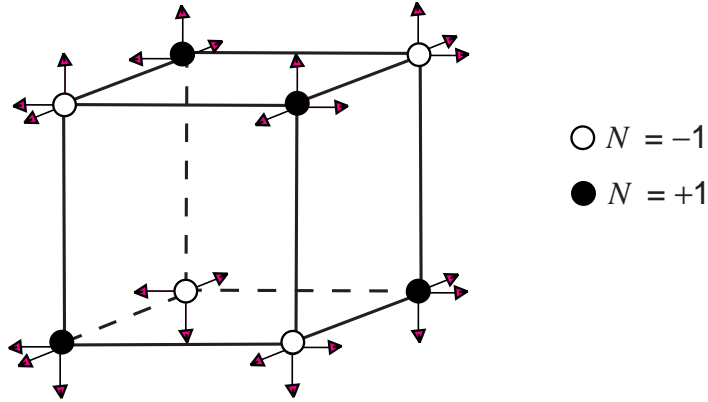}}
\label{Cube} 
  \caption{Sketch of topologically protected point nodes in momentum ${\bf}$ space with  topological charge $N=\pm 1$  in Eq.(\ref{MasslessTopInvariant3D})  in superconfuctors/superfluids 
  of $O(D_2)$ symmetry class \cite{VolovikGorkov1985}.  Chiral fermions emerge in the vicinity of each Fermi points. They  have identical emergent Minkowski metric, but different orientations of dreibein. The simplest realization of dreibein for each of eight chiral fermions is shown by arrows.
The vierbein orientations in $O(D_2)$ symmetry class superconductors are more complicated than in Figure:  one of the vectors in each vierbein is along the cube main diagonal, see Eq.(\ref{triad1}). All four pairs of left and right Weyl fermions have the same quantum numbers, while their triads  can be transformed to each other by rotations and reflection. They are analogous to generations of Standard Model fermions.
 }
\end{figure}

An example of the multiple splitting is provided by the model Hamiltonian for fermions  in superconfuctors/superfluids in the state which belongs to $O(D_2)$ symmetry class \cite{VolovikGorkov1985}. The  Bogoliubov - de Gennes (BdG)
Hamiltonian for fermions in this  spin-singlet $d$-wave superconductor has the form
 \begin{equation}
 H=\frac{1}{\sqrt{2}}(p^2-p_F^2)\tau_3 +  \frac{1}{2}(2p_x^2 -p_y^2-p_z^2)\tau_1 + \frac{\sqrt{3}}{2}(p_y^2-p_z^2)\tau_2 \,.
\label{Hamiltonian}
\end{equation}
At $p_F^2<0$ the energy spectrum is fully gapped, for $p_F^2=0$ the node in the spectrum appears at ${\bf p}=0$ which at $p_F^2>0$ splits into 8 Fermi points at the vertices of cube in momentum space
(see Fig. \ref{Cube}):
 \begin{equation}
{\bf p}^{(n)}=\frac{p_F}{\sqrt{3}}( \pm \hat{\bf x} \pm \hat{\bf y}   \pm \hat{\bf z} )~~,~~n=1, \ldots,8\,.
\label{Nodes}
\end{equation}
These nodes have topological charges $N=\pm 1$ in Eq.(\ref{MasslessTopInvariant3D}), and as a result,  close to each of  8 nodes the Hamiltonian is reduced to the Hamiltonian describing  Weyl fermions, 4 left and 4 right:
 \begin{equation}
 H^{(n)}={\bf e}_1^{(n)}\cdot ({\bf p}-{\bf p}^{(n)}) \tau_1  +{\bf e}_2^{(n)}\cdot  ({\bf p}-{\bf p}^{(n)}) \tau_2+ {\bf e}_3^{(n)}\cdot  ({\bf p}-{\bf p}^{(n)}) \tau_3\,.
\label{HamiltonianWeyl}
\end{equation}
Each Weyl fermion has its own triad (dreibein). Choosing for simplicity $p_F=\sqrt{3}$ one has
 \begin{eqnarray}
 {\bf e}_3^{(n)}= \sqrt{2}( \pm \hat{\bf x}   \pm \hat{\bf y}   \pm \hat{\bf z}) \,,
 \label{triad1}
 \\
  {\bf e}_1^{(n)}= \pm 2\hat{\bf x}   \mp \hat{\bf y}   \mp \hat{\bf z} \,,
  \label{triad2}
  \\
   {\bf e}_2^{(n)}= \sqrt{3}(  \pm \hat{\bf y}   \mp \hat{\bf z})
  \,.
\label{triad3}
\end{eqnarray}
All triads  can be transformed to each other by rotations and/or reflection. So  in this model one obtains four identical copies  of right and left relativistic Weyl fermions. They may be considered as analogs  of generations of Standard Model fermions, but with $n_g=4$. Different, but related mechanism 
for the origin of generations is suggested in \cite{Kaplan2011}.

 \section{Exotic fermions} 
\label{sec:Exotic_fermions}

In many systems (including condensed matter and relativistic quantum vacua), the Fermi points with elementary charges $N=\pm 1$ may merge together forming either the neutral point with $N=0$ or point with multiple $N$ (i.e. $|N| >1$ \cite{VolovikKonyshev1988}). In this case topology and symmetry become equally important, because it is the symmetry which may stabilize the degenerate node.
Example is provided by the Standard Model of particle physics, where 16 fermions of one generation have degenerate Dirac point at ${\bf p}=0$ with the trivial total topological charge $N=8-8=0$. In the symmetric phase of  Standard Model the nodes in the spectrum survive due to a discrete symmetry between the fermions and they disappear in the non-symmetric phase forming the fully gapped vacuum \cite{Volovik2003}. In case of degenerate Fermi point with $|N| >1$, situation is more diverse. Depending on symmetry, the interaction between fermionic flavors may lead to splitting of the multiple Fermi point to elementary Dirac points \cite{KlinkhamerVolovik2005a}; or gives rise to the essentially non-relativistic energy spectrum $E_\pm(p\rightarrow 0) \rightarrow \pm p^N$, which corresponds to different scaling for space and time in the infrared: ${\bf r}\rightarrow b {\bf r}$, $t\rightarrow b^N t$. The particular case of anisotropic scaling with  $N=3$ was suggested by Ho\v{r}ava for quantum gravity at short distances, the so-called  Ho\v{r}ava-Lifshitz gravity \cite{HoravaPRL2009,HoravaPRD2009,Horava2008}, while the anisotropic scaling in the infrared was suggested in Ref. \cite{Horava2010}. The topology of the multiple Dirac point provides another possible realization of anisotropic gravity, which is different from the scenario based on Lifshitz point in the theory of phase transitions  \cite{Lifshitz1941}. 

 \subsection{Dirac fermions with quadratic spectrum}
 \label{sec:Quadratic_spectrum} 

The non-linear spectrum arising near the Fermi point with $N=2$ has been discussed for different systems including graphene, double cuprate layer in high-$T_c$ superconductors, surface states of topological insulators and neutrino physics
 \cite{Volovik2001,Volovik2003,Volovik2007,Manes2007,Dietl-Piechon-Montambaux2008,Chong2008,Banerjee2009,Sun2010,Fu2011}. 
The spectrum of (quasi)particles in the vicinity of the doubly degenerate node depends on symmetry.    
Let us consider zeroes of co-dimension 2 -- point nodes in 2+1 systems. The node with topological charge $N=+ 2$ takes place in bilayered graphene.  According to general classification \cite{Horava2005},
the topology alone cannot protect zeroes of co-dimension 2:  the topological invariant takes place only in the presence of a symmetry. In particular, if we restrict consideration only to real (Majorana) fermions,
the nodes obey $Z_2$ topology  with summation law $1+1=0$ \cite{Horava2005}. To make  the multiple Fermi point possible we need an additional symmetry $K$ which extends the group $Z_2$ to the full group of integers $Z$. The relevant symmetry protected topological invariant is \cite{WenZee2002,Volovik2007,Manes2007,Beri2009}:
\begin{equation}
N= \frac{1}{4 \pi i}~
{\bf tr}\oint_C   dl 
~ K G(\omega=0,{\bf p})\partial_l G^{-1}(\omega=0,{\bf p}) ={\bf tr}\oint_C   dl 
~ K {\cal H}^{-1}({\bf p})\partial_l {\cal H}({\bf p})\,,
\label{MasslessTopInvariant1+1}
\end{equation}
where  $C$ is contour around the Dirac point in 2D momentum space $(p_x,p_y)$;  $K$ is the
relevant symmetry operator; $G$ is the Green's function matrix at zero frequency, which can be used as the effective Hamiltonian, ${\cal H}({\bf p})=G^{-1}(\omega=0,{\bf p})$; the operator $K$ commutes or anticommutes with the effective Hamiltonian.

Provided the symmetry $K$ is preserved and thus the summation law for $N$ takes place, one finds several scenarios of the behavior of the system with the total topological charge $N=+2$. 

(i) One may  have two fermions with the linear Dirac spectrum, with the nodes being at the same point of momentum space. This occurs if there is some special symmetry, such as the fundamental Lorentz invariance.

(ii) Exotic massless fermions emerge. In the 2D systems, these are gapless Dirac fermions with parabolic energy spectrum, which emerge at low energy:
\begin{equation}
E_\pm({\bf p})\approx \pm  {p^2} \,.
\label{WeylQuadratic}
\end{equation}
They are described by the following effective Hamiltonian
\begin{equation}
{\cal H}(p_x,p_y)=
\left( \begin{array}{cc}
0 &(p_x+ip_y)^2\\
(p_x-ip_y)^2&0
\end{array} \right)=(p_x^2-p_y^2)\sigma_1 - 2p_xp_y\sigma_2
\,,
\label{2x2}
\end{equation}
where  $\sigma_1$ and $\sigma_2$ are Pauli matrices.  The topological charge $N$ of the node at the point $p_x=p_y=0$ is given by the Eq.(\ref{MasslessTopInvariant1+1}), where the symmetry operator $K$  is represented by the Pauli matrix 
$\sigma_3$. With effective Hamiltonian (\ref{2x2}) one obtains that the node with the quadratic spectrum has the charge $N=2$.
In 3D systems, the corresponding fermions with the topological charge $N=2$ in Eq. (\ref{MasslessTopInvariant3D})  are the  semi-Dirac fermions, with linear dispersion in one direction and quadratic dispersion in the other
\cite{Volovik2001}:
  \begin{equation}
E_\pm({\bf p})\approx \pm \sqrt{c^2p_z^2 +{p_\perp^4}} \,.
\label{SemiDirac}
\end{equation}

\begin{figure}[t]
\centerline{\includegraphics[width=0.8\linewidth]{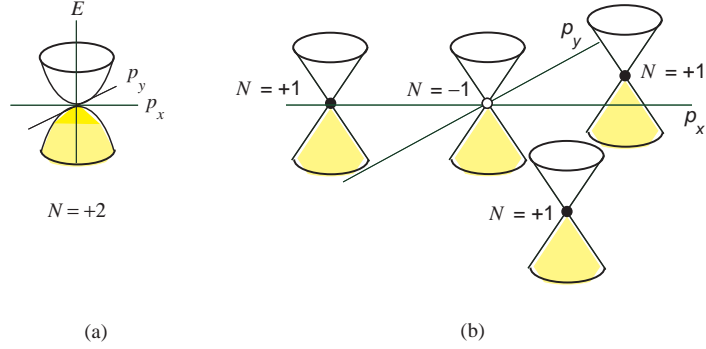}}
\medskip
\caption{Illustration of summation rule for momentum-space topological invariant.
Splitting of $N=2$ point with quadratic dispersion (a) into four Dirac points: $N=2=1+1+1-1$ (b).}
\label{TrigonalFig}
\end{figure}

(iii) The Weyl point with $N=+2$ may split either into two Weyl points each with $N=+1$ 
(see \cite{KlinkhamerVolovik2005a} for the relativistic 3+1 system) or into four Weyl points (three with $N=+1$ and one with $N=-1$, see Fig. \ref{TrigonalFig}). The effective Hamiltonian for the latter case is 
\cite{CannFalko2006,KoshinoAndo2006}: 
\begin{equation}
{\cal H}(p_x,p_y)=
\left( \begin{array}{cc}
0 &(p_x+ip_y)^2+s(p_x-ip_y)\\
(p_x-ip_y)^2+s(p_x+ip_y)&0
\end{array} \right)
\,.
\label{2x2warping}
\end{equation} 
The energy spectrum of this Hamiltonian has four Dirac point: the node at ${\bf p}=0$ has the topological charge $N=-1$, while three nodes at $p_x+ip_y=-se^{2\pi k i/3}$ with integer $k$ have charges $N=+1$ each, so that the summation rule $N=1+1+1-1=2$ does hold.

Note that in options (ii) and (iii) the (effective) Lorentz invariance of the Dirac point is violated.
This suggests that the topological mechanism of splitting of the Dirac point 
\cite{KlinkhamerVolovik2005a} or of the formation of the nonlinear dispersion \cite{Volovik2001} may lead to the spontaneous breaking of Lorentz invariance in the relativistic quantum vacuum, which in principle may occur in the neutrino sector of the quantum vacuum 
\cite{KlinkhamerVolovik2011,Klinkhamer2011a,Klinkhamer2011b}.

Let us consider the Fermi point with higher degeneracy, described by the symmetry protected topological invariant $N>2$.

 \subsection{Dirac fermions with cubic spectrum}
 \label{sec:Cubic_spectrum}

Let us consider first the case with $N=3$ \cite{HeikkilaVolovik2010}. Examples are three families of right-handed Weyl 2-component fermions in particle physics; three cuprate layers in high-$T_c$ superconductors; three graphene layers, etc. If the Fermi point is topologically protected, i.e. there is a conserved topological  invariant $N$, the node in the spectrum cannot disappear even in the presence of interaction, but it can split into $N$ nodes with elementary charge $N=1$. The splitting can be prevented if there is a symmetry in play, such as rotational symmetry. Here we provide an example of such symmetry.

\begin{figure}[h]
\centering
\includegraphics[width=8cm]{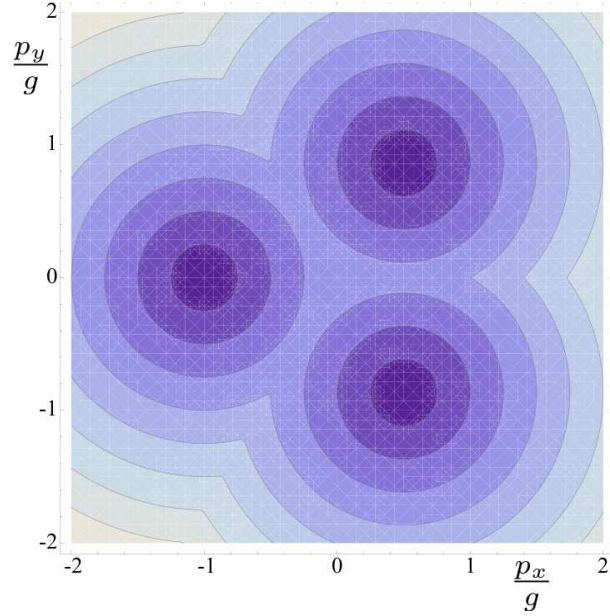}
\caption{
Splitting of Fermi point with $N=3$ into three Fermi points with $N=1$
in the model discussed in \cite{HeikkilaVolovik2010} .}
  \label{permuting_hamiltonian_spectrum_contour}
\end{figure}

We consider 3 species (families or flavors) of fermions, each of them being described by the invariant $N=+1$ in Eq.(\ref{MasslessTopInvariant1+1}) and an effective relativistic  Hamiltonian  emerging in the vicinity of the Fermi point
\begin{equation}
{\cal H}_0({\bf p})=\mbox{\boldmath$\sigma$}\cdot{\bf p}= \sigma_x p_x + \sigma_y p_y \,.
\label{NonInteractingSpecies}
\end{equation}
The matrix $K=\sigma_z$ anticommutes with the Hamiltonian. This supports 
the topologically protected node in spectrum, which is robust to interactions. The position of the node here is chosen at ${\bf p}=0$:
\begin{equation}
E^2=p^2 \,.
\label{NonInteractingSpectrum}
\end{equation}
The total topological charge of three nodes at ${\bf p}=0$ of three fermionic species is $N=+3$.
Let us now introduce matrix elements which mix the fermions. If these elements violate symmetry $K$, then the topological invariant cannot be constructed and point node will be destroyed, so let the matrix elements obey the symmetry $K$. For the general case of the matrix elements, but still obeying the symmetry $K$, the multiple node  will split into 3 or more elementary nodes, obeying the summation rule: $3=1+1+1=1+1+1+1-1= \cdots$ (see Fig. \ref{permuting_hamiltonian_spectrum_contour}).  
However, in the presence of some extra symmetry, which prevents splitting, the branch with the cubic spectrum emerges. 

 \begin{figure}[h]
\centering
\includegraphics[width=8cm]{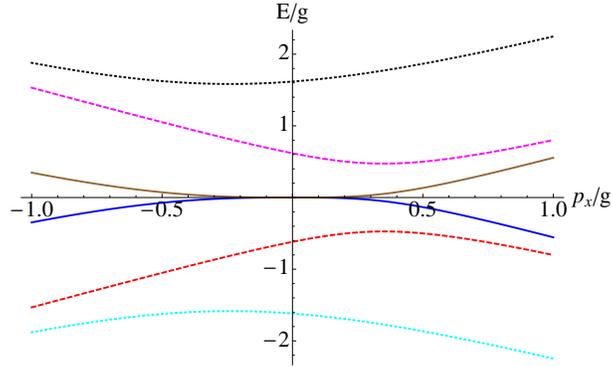}
\caption{
Spectrum of the Hamiltonian (\ref{3x3a}) showing cubic
  dispersion for the lowest two eigenvalues around the point
  $\mathbf{p}=0$. The spectrum has been calculated with equal
  coupling strengths $g_{12}=g_{13}=g_{23}=g$, but cubic spectrum 
  characterized by topological charge $N=3$ preserves for any nonzero values
  of coupling strengths. Spectrum is shown as function of $p_x$ at $p_y=0$. }
  \label{Cubic_spectrum}
\end{figure}
Example is provided by the following Hamiltonian \cite{HeikkilaVolovik2010} 
\begin{equation}
{\cal H}({\bf p})=
\left( \begin{array}{ccc}
\mbox{\boldmath$\sigma$}\cdot{\bf p} &g_{12}\sigma^+&g_{13}\sigma^+\\
g_{21}\sigma^-&\mbox{\boldmath$\sigma$}\cdot{\bf p} &g_{23}\sigma^+\\
g_{31}\sigma^-&g_{32}\sigma^-&\mbox{\boldmath$\sigma$}\cdot{\bf p}
\end{array} \right)
\,,
\label{3x3a}
\end{equation}
where $\sigma^\pm=\frac{1}{2}(\sigma_x\pm i\sigma_y)$ are ladder operators.
This Hamiltonian  anti-commutes with $K=\sigma_z$ and thus mixing preserves the topological charge
$N$ in (\ref{MasslessTopInvariant1+1}). 
At $p_x=p_y=0$ it is independent of the spin rotations up to a global phase of the coupling constants. Under spin rotation by angle $\theta$ all elements in the upper triangular matrix are multiplied by  $e^{i\theta}$, while all elements in the lower triangular matrix are multiplied by  $e^{-i\theta}$. This symmetry of triangular matrices  does not allow the multiple Fermi point to split  at $p=0$, as a result the gapless branch of spectrum in Fig. \ref{Cubic_spectrum}  has the cubic form at low energy, $E\rightarrow 0$, which corresponds to the topological charge $N=+3$:
\begin{equation}
E^2\approx \gamma_3^2 p^6~~,~~ \gamma_3=\frac{1}{|g_{12}||g_{23}|}\,.
 \label{CubicSpectrum}
\end{equation}
In the low-energy limit the spectrum in the vicinity of the multiple Fermi point (\ref{CubicSpectrum}) is symmetric under rotations. But in general the spectrum is not symmetric  as demonstrated in Figs. \ref{Cubic_spectrum}.
There is only the symmetry with respect to reflection, $(p_x,p_y) \rightarrow (p_x,-p_y)$. The rotational symmetry of spectrum (\ref{CubicSpectrum}) is an emergent phenomenon, which takes place only in the limit $p\rightarrow 0$.
 For trilayer graphene this spectrum has been discussed in Ref. \cite{Guinea2006}.

 \subsection{Dirac fermions with quartic and higher order spectrum}
 \label{sec:Quartic_spectrum}

\begin{figure}[h]
\centering
\includegraphics[width=8cm]{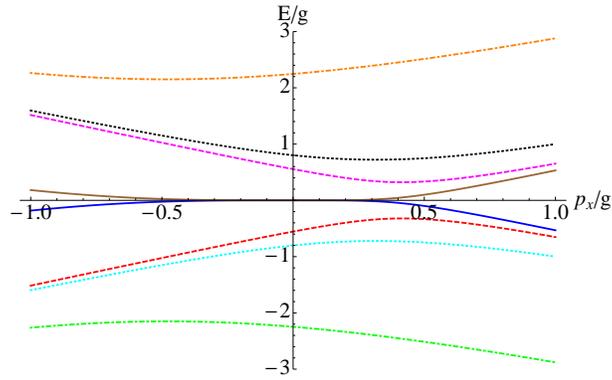}
\caption{
Spectrum of the Hamiltonian (\ref{4x4a}) showing quartic
  dispersion for the lowest two eigenvalues around the point
  $\mathbf{p}=0$. Different colors correspond to different
  eigenvalues, spectrum is shown as function of $p_x$ at $p_y=0$. The spectra have been calculated with equal coupling strengths $g_{12}=g_{13}=g_{23}=g_{14}=g_{24}=g_{34}=g$.}
\end{figure}

In case of four fermionic species, the mixing which does not produce splitting of the Fermi point is obtained by the same principle as in (\ref{3x3a}): all matrix elements above the main diagonal contain only  $\sigma^+$
(or $\sigma^-$):
\begin{equation}
{\cal H}({\bf p})=
\left( \begin{array}{cccc}
\mbox{\boldmath$\sigma$}\cdot{\bf p} &g_{12}\sigma^+&g_{13}\sigma^+&g_{14}\sigma^+\\
g_{21}\sigma^-&\mbox{\boldmath$\sigma$}\cdot{\bf p} &g_{23}\sigma^+&g_{24}\sigma^+\\
g_{31}\sigma^-&g_{32}\sigma^-&\mbox{\boldmath$\sigma$}\cdot{\bf p}&g_{34}\sigma^+\\
g_{41}\sigma^-&g_{42}\sigma^-&g_{42}\sigma^-&\mbox{\boldmath$\sigma$}\cdot{\bf p}
\end{array} \right)
\,.
\label{4x4a}
\end{equation}
Then again under spin rotation by angle $\theta$ all elements in the upper triangular matrix are multiplied by  $e^{i\theta}$, while all elements in the lower triangular matrix are multiplied by  $e^{-i\theta}$. As a result the multiple Fermi point with $N=4$ is preserved giving rise to the quartic spectrum in vicinity of the Fermi point:
\begin{equation}
E^2=\gamma_4^2 p^8~~,~~ \gamma_4=\frac{1}{|g_{12}||g_{23}||g_{34}|}\,.
 \label{QuarticSpectrum}
\end{equation}
For the tetralayer graphene this spectrum was suggested in Ref. \cite{Mak2010}.
Again the rotational symmetry emerges new the Fermi point, but spectrum is not symmetric under permutations. In fact, there is no such $4\times 4$ matrix that would consist of couplings described by ladder operators and which would be symmetric under permutations.

In general the Fermi point with arbitrary $N$ may give rise to the spectrum
\begin{equation}
E^2=\gamma_N^2 p^{2N}\,.
 \label{NSpectrum}
\end{equation}
Such spectrum emerges in multilayered graphene   \cite{Manes2007,Castro2009}.
 The discussed symmetry of matrix elements $g_{mn}$ extended to  $2N\times 2N$ matrix gives 
(\ref{NSpectrum}) with the prefactor
\begin{equation}
\gamma_N=\frac{1}{|g_{12}||g_{23}|\ldots |g_{N-1,N}|}\,.
 \label{NSpectrum2}
\end{equation}
Violation of this symmetry may lead to splitting of the multiple Fermi point  into
  $N$ elementary Fermi points -- Dirac points with $N=1$ and `relativistic' spectrum
  $E^2 \propto p^{2}$. 
  
The effective Hamiltonian describing fermions in the vicinity of multiple Fermi point is
$H=\sigma^-p_+^N + \sigma^+p_-^N$, see  \cite{Manes2007,Cortijo2011}. An example of the effective Hamiltonian describing the multiple Fermi point with topological charge $N$ in 3+1 systems is \cite{VolovikKonyshev1988}
\begin{equation}
H=\sigma_z p_z + \sigma^-p_+^N + \sigma^+p_-^N\,.
 \label{3+1}
\end{equation}
This Hamiltonian has the spectrum $E^2=p_z^2 + p_\perp^{2N}$,  which has linear dispersion in one direction and non-linear dispersion in the others.

This is also applicable to the vacuum of particle physics. The Lorentz symmetry prohibits both the splitting of the Dirac points and the non-linear non-relativistic spectrum. Situation changes if the Lorentz symmetry is viewed as an emergent phenomenon, which arises near the Dirac point (Fermi points with $N=\pm 1$). In this case both both splitting of Dirac points  and formation of non-linear non-relativistic spectrum in the vicinity of the multiple Fermi point are possible, an the choice depends on symmetry. In both cases the mixing of fermions violates the effective Lorentz symmetry in the low-energy corner. This phenomenon,  called the reentrant violation of special relativity \cite{Volovik2001}, has been discussed for $N_F = 3$ fermion families in relation to neutrino oscillations
\cite{Klinkhamer2006}. Influence of possible discrete flavor symmetries on neutrino mixing has been reviewed in Ref. \cite{Altarelli2010}.

\section{Flat bands in topological media}
 \label{sec:Flat_bands}

\begin{figure}[h]
\centering
\includegraphics[width=0.4\linewidth]{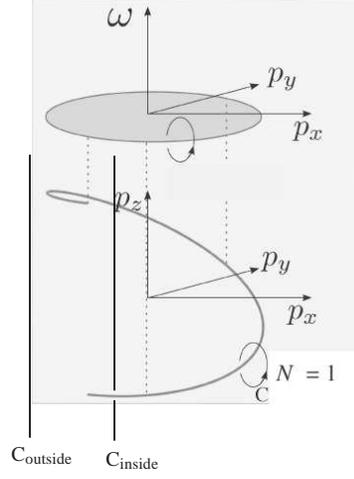}
\caption{Nodal spiral generates topologically protecetd flat band on the surface. Projection of spiral on the surface determines boundary of flat band. At each $(p_x,p_y)$ except the boundary of circle $p_x^2+p_y^2=t^2$
the system represents the 1D gapped state (insulator). At each $(p_x,p_y)$ inside the circle, the
insulator is topological being described by non-zero topological invariant (\ref{InvariantForLine2}) and thus one has gapless edge state. The manifold of these zero-energe edge state inside the circle
forms the flat band  found in Eq.(\ref{Flat_Band}). }
\label{fig:spiral}
\end{figure}

Let us extend the results of the Sec. \ref{sec:Quartic_spectrum}   to the limit case of infinite number of layers, $N\rightarrow \infty$ \cite{HeikkilaVolovik2011}. If the layers are equivalent and interact only via nearest neighbor couplings, i.e. the nonzero matrix elements are
\begin{equation}
g_{12}=g_{23}=\ldots g_{N-1,N}\equiv t\,,
 \label{Hopping}
\end{equation} 
the low-energy spectrum becomes 
\begin{equation}
E=t \left(\frac{p}{t}\right)^{N}\,.
 \label{NSpectrum_equal_g}
\end{equation}
In the $N\rightarrow \infty$ limit one obtains that in the lowest energy branch all the fermions within circumference $|{\bf p}| = t$ have exactly zero energy.
\begin{equation}
E\left(N\rightarrow \infty, |{\bf p}|< t\right) =0\,.
 \label{Flat_Band}
\end{equation}
Here we consider the topological origin of the flat band. When the number of layers $N\rightarrow \infty$,
the quasi two dimensional system transforms to the $D=3$ system. We find that the flat band emerges in the first and the last layers, i.e. on the boundaries of the 3D system. Its formation is accompanied by the  simultaneous formation
of the line of node (zeroes of co-dimension 2) in the bulk material. These two objects, surface flat band
and line of zeroes in bulk, are connected via a special kind of the topological bulk-surface correspondence: projection of the nodal line to the surface determines the boundary of the flat band
(Fig. \ref {fig:spiral}).

\subsection{Topological origin of surface flat band}
 \label{sec:Origin_Flat_bands} 
 
To understand the topological origin of this branch and its structure
let us consider the spectrum in the continuous limit.
The effective Hamiltonian in the 3-dimensional bulk system which emerges in the limit 
of infinite number of layers is the following $2\times 2$ matrix
\begin{equation}
H=
\left( \begin{array}{cc}
0 &f\\
f^*&0
\end{array} \right)
~~,~~f=p_x-ip_y - t e^{-ia p_z}  \,.
\label{ContinuousH}
\end{equation}
Here $t$ is the magnitude of the hopping matrix element between the layers in Eq.(\ref{Hopping}). The energy spectrum of the bulk system 
\begin{equation}
E^2= [p_x - t\cos  (ap_z)]^2+  [p_y +t\sin  (ap_z)]^2 \,,
\label{ContinuousE}
\end{equation}
 has zeroes on line   (see Fig.~\ref{fig:spiral}):
 \begin{equation}
 p_x =t\cos  (ap_z) ~~,~~ p_y =  -t \sin  (ap_z) \,,
\label{NodalLine}
\end{equation}
which forms a spiral, the projection of this spiral on the plane $p_z={\rm const}$ being the circle 
 ${\bf p}_\perp^2\equiv p_x^2+p_y^2= t^2$.
This nodal line is topologically protected by the same topological invariant as in 
Eq.(\ref{MasslessTopInvariant1+1})
  \begin{equation} 
N= {1\over 4\pi i} ~{\rm tr} ~\oint_C dl ~ \sigma_z H^{-1}\nabla_l H \,, 
\label{InvariantForLine}
\end{equation}
where the integral is now along the loop $C$ around the nodal line in momentum space, see Fig. \ref{fig:spiral}. 
The  winding number around the element of the nodal line is $N=1$.

\begin{figure}[h]
\centering
\includegraphics[width=8cm]{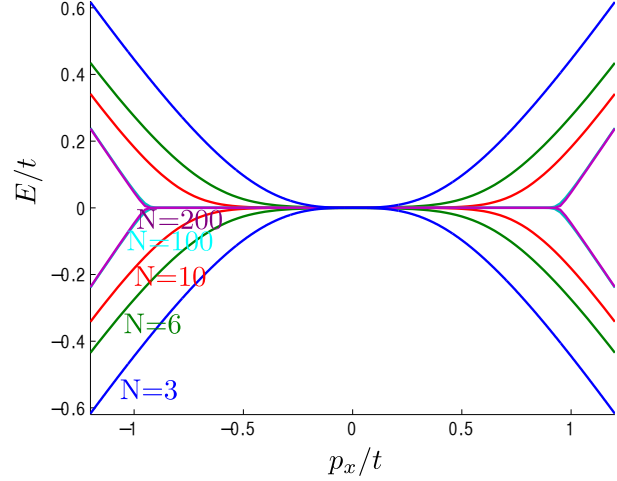}
\caption{Formation of the surface flat band. When  the number $N$ of layers increases, the  dispersionless band evolves from the   gapless branch of the spectrum, which has the form $E=\pm |{\bf p}_\perp|^N$ in the vicinity of multiple Dirac point. The spectrum is shown as a  function of $p_x$ for $p_y=0$. The curves for $N=100$ and $N=200$ are almost on top of each other. Asymptotically  the spectrum $E=\pm |{\bf p}_\perp|^N$ transforms to the dispersionless band within the projection of the nodal line to the surface. }
    \label{spectrum_multiple_Dirac}
\end{figure}

Let us consider now the momentum ${\bf p}_\perp$ as a parameter of the 1D system, then for
$|{\bf p}_\perp|\neq t$ the system represents the fully gapped system -- 1D insulators.
This insulator can be described by the same invariant as in Eq.(\ref{InvariantForLine}) with 
 the contour of integration chosen parallel to $p_z$, i.e. along the 1D Brillouin zone at fixed 
 ${\bf p}_\perp$ (due to periodic boundary conditions, the points $p_z=\pm \pi/a$ are equivalent and the contour of integrations forms the closed loop):
\begin{equation} 
N({\bf p}_\perp)= {1\over 4\pi i} ~{\rm tr} ~\int_{-\pi/a}^{+\pi/a} dp_z ~ \sigma_z H^{-1}\nabla_{p_z} H\,.
\label{InvariantForLine2}
\end{equation}
For  $|{\bf p}_\perp|<t$ the 1D insulator is topological, since $N({|\bf p}_\perp|<t)=1$, while for $|{\bf p}_\perp|> t$ one has $N_1({\bf p}_\perp)=0$ and the 1D insulator is the trivial band insulator. The line $|{\bf p}_\perp|=t$ thus marks the topological quantum phase transition between the topological and non-topological 1D insulators.

Topological invariant $N({\bf p}_\perp)$ in (\ref{InvariantForLine2}) determines the property of the surface bound states of the 1D system at each ${\bf p}_\perp$. Due to the bulk-edge correspondence, the topological 1D insulator must have the surface state with exactly zero energy. Since such states exist for any parameter within the circle  $|{\bf p}_\perp|=t$, one obtains the flat band of surface states with exactly zero energy, $E(|{\bf p}_\perp|<t)=0$, which is protected by topology. This is the origin of the unusual branch of spectrum in Eq.(\ref{Flat_Band}): it represents the band of topologically protected surface states with exactly zero energy. Such states do not exist for parameters $|{\bf p}_\perp|>t$, for which the 1D insulator is non-topological.

The zero energy bound states on the surface  of the system can be obtained directly from the Hamiltonian:
\begin{equation}
\hat H=\sigma_x (p_x -t\cos (a\hat p_z)) + \sigma_y (p_y + t\sin (a\hat p_z)) 
~~,~~
\hat p_z=-i\partial_z~~,~~ z<0\,.
\label{ContinuousHamilton}
\end{equation}
We assumed that the system occupies the half-space $z<0$ with the boundary at $z=0$.
This Hamiltonian has the bound state with exactly zero energy, $E({\bf p}_\perp)=0$,  for any $|{\bf p}_\perp|<t$, with the eigenfunction concentrated near the surface:
 \begin{equation}
\Psi \propto \left( \begin{array}{cc}
0\\
1
\end{array} \right)
(p_x-ip_y) \exp{\frac{z \ln(t/(p_x+i p_y))}{a}}~~,~~|{\bf p}_\perp|< t \,.
\label{SurfaceWaveFunction}
\end{equation}
 The normalizable wave functions with zero energy exist only  for ${\bf p}_\perp$  within the circle $|{\bf p}_\perp|\leq t$, i.e. the surface flat band  is bounded by the projection of the nodal spiral onto the surface. Such correspondence between the flat band on the surface and lines of zeroes in the bulk has been also found in Ref.  \cite{SchnyderRyu2010} for superconductors without inversion symmetry.

\begin{figure}[h]
\centering
\includegraphics[width=8cm]{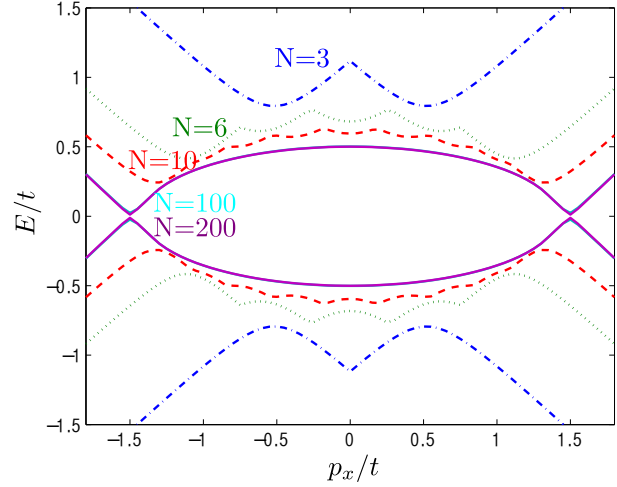}
\caption{ Formation of the nodal line from the evolution of the gapped branch of the spectrum of the multilayered system, when the number $N$ of layers increases.  
The spectrum is shown as function of $p_x$ for $p_y=0$. 
The curves for $N=100$ and $N=200$ lie almost on top of each other, indicating the bulk limit. Asymptotically the nodal line  $p_x =t\cos  (ap_z)$, $ p_y =  -t \sin  (ap_z)$ is formed (two points on this line are shown, which correspond to $p_y=0$).}
    \label{spectrum1}
\end{figure}

\subsection{Dimensional crossover in topological matter: Formation of the flat band in multilayered system}
\label{flatbandformation}

\begin{figure}[h]
\centering
\includegraphics[width=8cm]{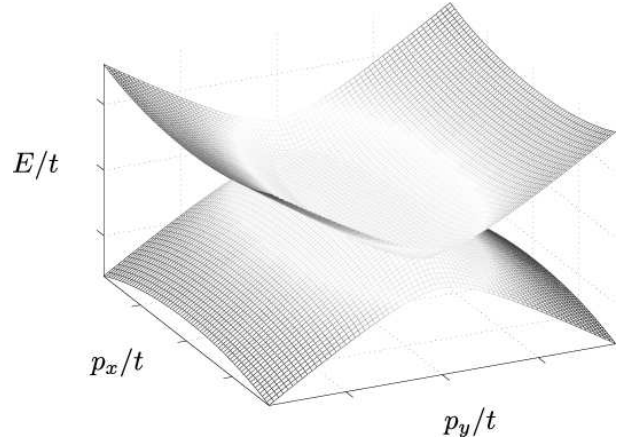}
\caption{The lowest energy states for different $p_x$ and $p_y$ and arbitrary $p_z$ in the bulk limit. The flat band of surface states is formed in the region  $p_x^2 +p_y^2 <t^2$. Outside this region only the bulk states exist. In a simple model considered here flat band comes from the degenerate Dirac point with nonlinear dispersion. However, this is not necessary condition: the flat band emerges whenever 
the nodal line appears in the bulk.}
\label{lowenspectrummesh}
\end{figure}

\begin{figure}[h]
\centering
\includegraphics[width=8cm]{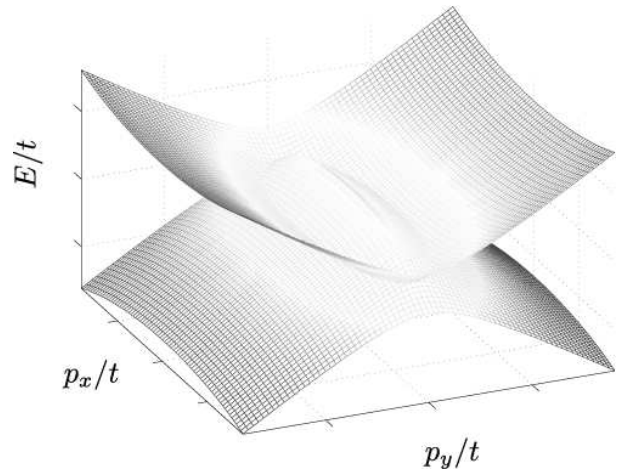}
\caption{Result of the transformation of the gapped states in the process of dimensional crossover. They form the nodal line in bulk $p_x =t \cos  (ap_z)$, $ p_y = -t\sin  (ap_z)$ whose projection   to the $(p_x,p_y)$-plane is shown.}
    \label{spectrummesh}
\end{figure}

The discrete model with finite number $N$ of layers has been considered in 
Ref. \cite{HeikkilaVolovik2011}. It is described by the $2N\times 2N$ Hamiltonian with the nearest neighbor interaction between the layers in the form:
\begin{equation}
H_{ij}({\bf p}_\perp)=
\mbox{\boldmath$\sigma$}\cdot{\bf p}_\perp \delta_{ij} - t \sigma^+ \delta_{i,j+1} -t\sigma^-\delta_{i,j-1}
~~,~~
 1\leq i\leq N~~,~~{\bf p}_\perp=(p_x,p_y)\,.
\label{DiscreteHamiltonian} 
\end{equation}
In the continuous limit of infinite number of layers (\ref{DiscreteHamiltonian}) transforms to 
(\ref{ContinuousH}) with the nodal line in the spectrum, while for finite $N$ the spectrum contains the Dirac point with multiple topological charge equal to $N$. 
Figures (\ref{spectrum_multiple_Dirac}) - (\ref{spectrummesh}) demonstrate how this crossover
from 2D to 3D occurs.  When  the number $N$ of layers increases, the  dispersionless surface band evolves from the   gapless branch of the spectrum $E=\pm |{\bf p}_\perp|^N$. Simultaneously, the gapped branches of the spectrum of the finite-$N$ system give rise to the nodal line in bulk.
This scenario of formation of the surface flat band takes place if the symmetry does not allow the splitting of the multiple Dirac point, and it continuously evolves  to the dispersionless spectrum.

\begin{figure}[h]
\centering
\includegraphics[width=8cm]{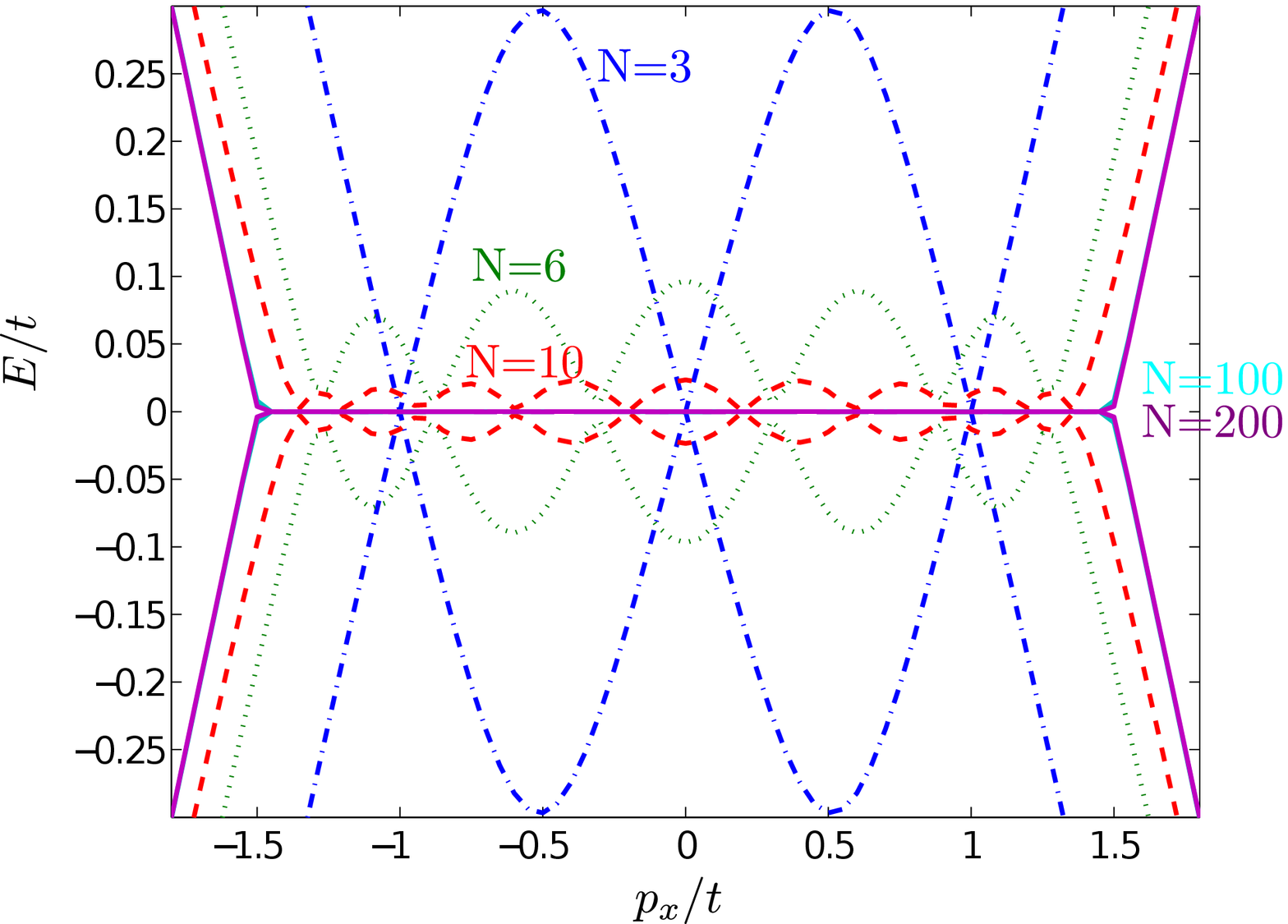}
\caption{Evolution of the spectrum in a  different model, in which the multiple Dirac point is absent,
but the surface flat band is formed together with the formation of the nodal line in bulk}
    \label{spectrum2}
\end{figure}

If this symmetry is absent, but the symmetry supporting the topological charge persists, the scenario 
of the flat band formation is different but still is governed by topology. Fig. (\ref{spectrum2}) demonstrates
the formation of the flat band in this situation. The reason, why the flat band emerges is the formation of the topologically protected nodal line in bulk. The projection of the nodal line on a surface
gives the boundary of the flat band emerging on this surface. This is the realization of  the bulk-surface correspondence in systems with the nodal lines in bulk. The other examples can be found in Ref. \cite{HeikkilaKopninVolovik2011}.

\section{Anisotropic scaling and Ho\v{r}ava gravity}
 \label{sec:HoravaGravity} 

We know that in the vicinity of the Weyl point with elementary topological charge $N=+1$ or $N=-1$ the quantum electrodynamics and gravity emerge as effective fields.
Here we discuss what kind quantum electrodynamics emerges near the node with higher $N$. This is the   quantum electrodynamics with anisotropic scaling, in which as distinct from isotropic relativistic QED the electric and magnetic fields obey different scaling laws. The analogous anisotropic scaling
has been discussed for quantum gravity at short distances.

\subsection{Effective theory near the degenerate Dirac point}
 \label{sec:EffectiveTheory} 

Let us consider again the 2+1 system with the multiple Dirac point and write the following effective Hamiltonian:
\begin{equation} 
{\cal H}_N= \frac{\sigma^x+i\sigma^y}{2}\left(({\bf e}_1+i{\bf e}_2)\cdot({\bf p}- e{\bf A}) \right)^{|N|}
+
 \frac{\sigma^x-i\sigma^y}{2}\left(({\bf e}_1-i{\bf e}_2)\cdot({\bf p}- e{\bf A})\right)^{|N|} \,.
\label{FermionHamiltonianN}
\end{equation}
For a single layer ($N=1$) the Hamiltonian (\ref{FermionHamiltonianN}) is reduced to the conventional Weyl-Dirac Hamiltonian for massless particles in 2+1 dimension:
\begin{equation} 
{\cal H}_{N=1}=\sigma^x {\bf e}_1 \cdot({\bf p}- e{\bf A})+ \sigma^y {\bf e}_2 \cdot({\bf p}- e{\bf A})= e_a^i \sigma^a(p_i-eA_i)~, ~a=(1,2)\,.
\label{FermionHamiltonianN=1}
\end{equation}
Here the vectors ${\bf e}_1$ and ${\bf e}_2$ play the role of zweibein (in the ground state they are mutually orthogonal). The vector ${\bf A}$  is either the vector potential of the effective electromagnetic field which comes from the shifts of the node, or the real electromagnetic field, as it takes place for electrons in graphene.  
The whole dreibein $e^\mu_a$ with $a=(1,2,3)$ and $\mu=(0,1,2)$ emerges for the Green's function and gives the effective metric 2+1 metric as a secondary object:
\begin{equation} 
g^{\mu\nu}=\eta^{ab} e^\mu_a  e^\nu_b ~, 
\label{MetricFromDreibein}
\end{equation}

For general $|N|$ situation is somewhat different. While the action for the relativistic fermions is invariant under rescaling ${\bf r} = b {\bf r}'$, $t = b  t'$, the action for the fermions living in the vicinity of the multiple Dirac point is invariant under anisotropic rescaling ${\bf r} = b {\bf r}'$, $t = b^{|N|}  t'$. The anisotropic scaling is in the basis of the Ho\v{r}ava gravity,  which is described by the space components of metric are separated from the time component and have different scaling laws \cite{HoravaPRL2009,HoravaPRD2009,Horava2008,Horava2010}. The square of Hamiltonian (\ref{FermionHamiltonianN}) gives the space metric in terms of zweibein:
\begin{equation} 
{\cal H}_N^2=E_N^2= \left(g^{ij}(p_i-eA_i)(p_j-eA_j)\right)^{|N|}~~,~~g^{ij}=e_1^i e_1^j+e_2^i e_2^j\,.
\label{EnergyN}
\end{equation}

\subsection{Effective electromagnetic  action}
 \label{sec:ElectromagneticAction}

If the effective action for fields $e^\mu_a$ and $A_\mu$ is obtained by integration over fermions
in the vicinity of the multiple Dirac point, this bosonic action (actually the terms in action which mostly come from these fermions) inherits the corresponding conformal symmetry of the massless fermions.
For anisotropic scaling the conformal invariance means invariance under $g^{ik} \rightarrow b^2 g^{ik}$ and $g^{00} \rightarrow b^{2|N|}g^{00}$; $\sqrt{-g} \rightarrow b^{-(|N|+D)}$ (where $D$ is space dimension); while $g^{0i}$ is not considered. 

\subsubsection{Single layered graphene and relativistic fields}
 \label{sec: SingleGraphene} 

For $D\neq 2$ the spectrum of multiple Fermi point becomes more complicated, and in general is not isotropic, see Eq.(\ref{3+1}). For general $D$ the spectrum is isotropic only 
for  $|N|=1$, where one obtains effective relativistic massless $D$+1 quantum electrodynamics,
which is Lorentz invariant. This implies the following nonlinear action
\begin{equation} 
S_{em}(|N|=1,D)=\int d^Dx dt \left[ B^2-E^2\right]^\frac{D+1}{4}\,.
\label{EMaction21}
\end{equation}
For $D=3$ the action is proportional to $(B^2-E^2) \ln (B^2-E^2)$, and is imaginary at $B^2<E^2$ giving rise to Schwinger pair production in massless quantum electrodynamics. 
The similar imaginary action tales place for $B^2<E^2$ for $D\neq 3$.
For example, for a single layer graphene ($D=2$, $|N|=1$) reproduces the relativistic 2+1 QED which gives rise to Lagrangian $(B^2-E^2)^{3/4}$
\cite{AndersenHaugset1995} with the running coupling constant 
$1/\alpha=\sqrt{2}\zeta(3/2)/8\pi^2$.  The action is imaginary at $B^2<E^2$ which corresponds to Schwinger pair production with the rate $E^{3/2}$ at $B=0$.

\subsubsection{Bilayer graphene}
 \label{sec: BilayerGraphene} 

For  bilayered graphene, assuming the quadratic dispersion $|N|=D=2$, the expected conformal invariant Heisenberg-Euler action for 
the constant in space and time electromagnetic field, which is
obtained by the integration over the 2+1 fermions with quadratic
dispersion, is the function of the scale invariant  combination $\mu$
\cite{KatsnelsonVolovik2012,KatsnelsonVolovikZubkov2012}:
 \begin{equation}
S\sim \int d^2x dt  ~ B^2 g(\mu)~~,~~ \mu=\frac{E^2}{B^3}\,.
\label{EffectiveActionGeneral}
\end{equation}
The asymptotical
behavior in two limit cases, $g(\mu \rightarrow 0)\sim const$ and $g(\mu
\rightarrow \infty)\sim \mu^{2/3}$, gives the effective actions
for the constant in space and time magnetic and electric fields:
 \begin{equation}
S_{B}=a \int d^2x dt  ~B^2  ~~,~~ S_{E}=  (b+ic) \int d^2x dt   E^{4/3} \,.
\label{BandE}
\end{equation}
The  parameter $a$ is the logarithmic coupling constant; the parameter $b$ describes the vacuum
electric polarization; and the parameter $c$ describes the
instability of the vacuum with respect to the Schwinger pair
production in the electric field, which leads to the imaginary
part of the action. 
The action also contains  the linear non-local  term 
\begin{equation} 
S_{non-local}(|N|=D=2)=\int d^2x dt \sqrt{-g} g^{00}g^{kn} F_{0k} \frac{1}{g^{ip}\nabla_i \nabla_p} F_{0n} 
\,,
\label{EMaction}
\end{equation}
which corresponds to the polarization operator
\begin{equation} 
\Pi_{00} \propto \frac{k^2}{\sqrt{k^4-\omega^2}}
~~,~~|N|=D=2\,.
\label{Polarization2}
\end{equation}

\subsubsection{D=2 systems with nodes with topological charge $N$}
 \label{sec: MultilayeredGraphene}

In general case of a 2D system with $N$-th order touching point in spectrum  (a kind of multilayered graphene) the Heisenberg-Euler action  contains among the other terms the following nonlinear terms in the actions for magnetic and electric fields \cite{KatsnelsonVolovik2012}:
\begin{equation} 
S_B(N,D=2) \sim \int d^2x dt  B^\frac{2+|N|}{2}~~
,~~S_E(N,D=2) \sim \int d^2x dt (-E^2)^{\frac{2+|N|}{2(1+|N|)}}\,,
\label{EMactionGeneral}
\end{equation}
where imaginary part of the action is responsible for the pair production in electric field
\cite{KatsnelsonVolovik2012,Zubkov2012}, while the linear action of the type (\ref{EMaction})
corresponds to the polarization operator
\begin{equation} 
\Pi_{00} \propto \frac{k^2}{\sqrt{k^{2N}-\omega^2}}
~~,~~D=2\,.
\label{PolarizationN}
\end{equation}

\subsubsection{Effective  action for gravity}
 \label{sec:GravityAction} 

The expected action for gravitational field is
\begin{equation} 
S_{grav}=\int d^2x dt \sqrt{-g} \left[ K_1 R^2 +K_2 g^{ik}g^{mn} g^{00}\partial_t g_{im} \partial_t g_{kn}+ \ldots\right]\,,
\label{Gaction}
\end{equation}
where $K_1, K_2,\ldots$ are dimensionless quantities, and the other terms of that type are implied.

\section{Fully gapped topological media}

Examples of the fully gapped topological media are the so-called
 topological band  insulators in crystals  \cite{HasanKane2010}.
Examples are Bi$_2$Se$_3$, Bi$_2$Te$_3$ and Sb$_2$Te$_3$ compounds which are predicted to be 3+1  topological insulators \cite{Kane2005}.
But the first discussion of the 3+1 topological insulators can be found in Refs. \cite{Volkov1981,VolkovPankratov1985}. The main feature of such materials is that they are insulators in bulk, where electron spectrum  has a gap, but there are  2+1 gapless edge states of electrons on the surface or  at the
interface between topologically different bulk states as discussed in Ref.  \cite{VolkovPankratov1985}. The similar properties 
are shared by the fully gapped 3D topological superfluids and superconductors. The spin triplet $p$-wave superfluid $^3$He-B represents the fully gapped superfluid  with nontrivial topology. It has 2+1 gapless quasiparticles living at interfaces between vacua with different values of the topological invariant describing the bulk states of $^3$He-B \cite{SalomaaVolovik1988,Volovik2009}.   The quantum vacuum of
Standard Model below the electroweak transition, i.e. in its massive phase, also shares the properties 
of the topological insulators and gapped topological superfluids and is actually the relativistic counterpart 
of $^3$He-B \cite{Volovik2010a}.

Examples of the  2+1  topological fully gapped systems are provided by the 
films of superfluid $^3$He-A with broken time reversal symmetry \cite{VolovikYakovenko1989,Volovik1992b} and by the  planar phase which is time reversal invariant
\cite{VolovikYakovenko1989,Volovik1992b}. The topological invariants for 2+1 vacua give rise to  quantization of the Hall and spin-Hall conducticity in these films in the
absence of external magnetic field (the so-called intrinsic qauntum and spin-quantum Hall effects) \cite{VolovikYakovenko1989,SQHE}.

 \subsection{2+1 fully gapped vacua}

\subsubsection{$^3$He-A film: 2+1 chiral superfluid}

The gapped (nodeless) ground states (vacua) in 2+1 systems are characterized by the invariant obtained by dimensional reduction
from the topological invariant describing the nodes of co-dimension 3. The invariant $N$
for the
Fermi point  in (\ref{MasslessTopInvariant3D} is the integral over 
the 3D surface $\sigma$ around the singularity. The invariant describing the 2D insulator
or nodeless supefluid is the integral over the whole (2+1)-dimensional momentum-frequency space $(p_x.p_y,\omega)$:
\begin{equation}
N = \frac{e_{ijk}}{24\pi^2}~
{\bf tr}\left[  \int    d^2p d\omega
~G\partial_{p_i} G^{-1}
G\partial_{p_j} G^{-1}G\partial_{p_k}  G^{-1}\right].
\label{N2+1}
\end{equation}

\begin{figure}[t]
\centerline{\includegraphics[width=0.7\linewidth]{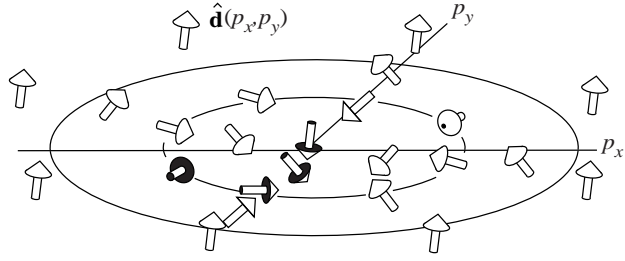}}
\caption{Skyrmion in ${\bf p}$-space with momentum space topological
charge
$N=-1$ in (\ref{N2+1d}). It describes topologically non-trivial vacua in 2+1
systems with a fully gapped non-singular Green's function. Vacua with nonzero $N$ 
have topologically protected gapless edge states. The nonzero topological charge leads also to quantization of Hall and spin Hall conductance.}
\label{PSpaceSkyrmionsFig}
\end{figure}

This equation (\ref{N2+1})
was introduced in relativistic $2+1$ theories \cite{So1985,IshikawaMatsuyama1986,IshikawaMatsuyama1987}
and for the film of $^3$He-A in condensed matter \cite{Volovik1988,VolovikYakovenko1989}, where it was inspired by
the dimensional reduction from the Fermi point, see \cite{Volovik1992b}.
In simple case of the $2\times 2$ matrix, the Green's function can be expressed in terms
of the three-dimensional vector  ${\bf d}(p_x,p_y)$, 
\begin{equation}
 G^{-1}(\omega,p_x.p_y)=i\omega -H ~~,~~ H={\mbox{\boldmath$\tau$}} \cdot {\bf d}(p_x,p_y)\,,
\label{d-vector}
\end{equation}
where  $H$ is the Bogoliubov - de Gennes Hamiltonian for fermions in  $p$-wave superfluids, and 
$\mbox{\boldmath$\tau$}$ are the Pauli matrices.  Example of the   ${\bf d}$-vector configuration, which corresponds to the  topologically nontrivial vacuum is presented in  Fig. \ref{PSpaceSkyrmionsFig}. This is the momemtum-space analog of the topological object in real space -- skyrmion.  In real space, skyrmions are described by the relative homotopy groups \cite{MineevVolovik1978}; they have been investigated in detail both theoretically and experimentally in the A phase of $^3$He, see  Sec. 16.2 in \cite{Volovik2003} and the review paper 
\cite{SalomaaVolovik1987}. 

For the Green's function in (\ref{d-vector}) the winding number of the momentum-space skyrmion in Eq.(\ref{N2+1}) is reduced to
 \cite{Volovik1988}
\begin{equation}
N= \frac{1}{4\pi} ~
   \int    d^2p  
~\hat{\bf d}\cdot \left(\frac{\partial \hat{\bf d}}{\partial p_x}
\times \frac{\partial \hat{\bf d}}{\partial p_y} \right) \,,
\label{N2+1d}
\end{equation}
where $\hat{\bf d}={\bf d}/|{\bf d}|$ is unit vector.
For a single layer of the $^3$He-A film and for one spin projection, the simplified Bogoliubov - de Gennes Hamiltonian has the form:
\begin{equation}
H={\mbox{\boldmath$\tau$}} \cdot {\bf d}({\bf p}) =
 i \omega +  \tau_3\left(\frac{p_x^2+p_y^2}{2m} -\mu\right) + \tau_1 p_x  + \tau_2 p_y~,
\label{3He-A_film}
\end{equation}
For $\mu>0$ the topological charge in Eq.(\ref{N2+1d}) or in Eq.(\ref{N2+1}) is $N=1$, and
the  $\hat{\bf d}({\bf p})$ field forms the skyrmion in momentum space of the type shown in 
  Fig. \ref{PSpaceSkyrmionsFig}.
For $\mu<0$ the topological charge is trivial, $N=0$. That is why at $\mu=0$ there is a topological quantum phase transition between the topological superfluid at $\mu>0$ and non-topological superfluid at $\mu<0$ \cite{Volovik1992b}.

\begin{figure}[t]
\centerline{\includegraphics[width=0.7\linewidth]{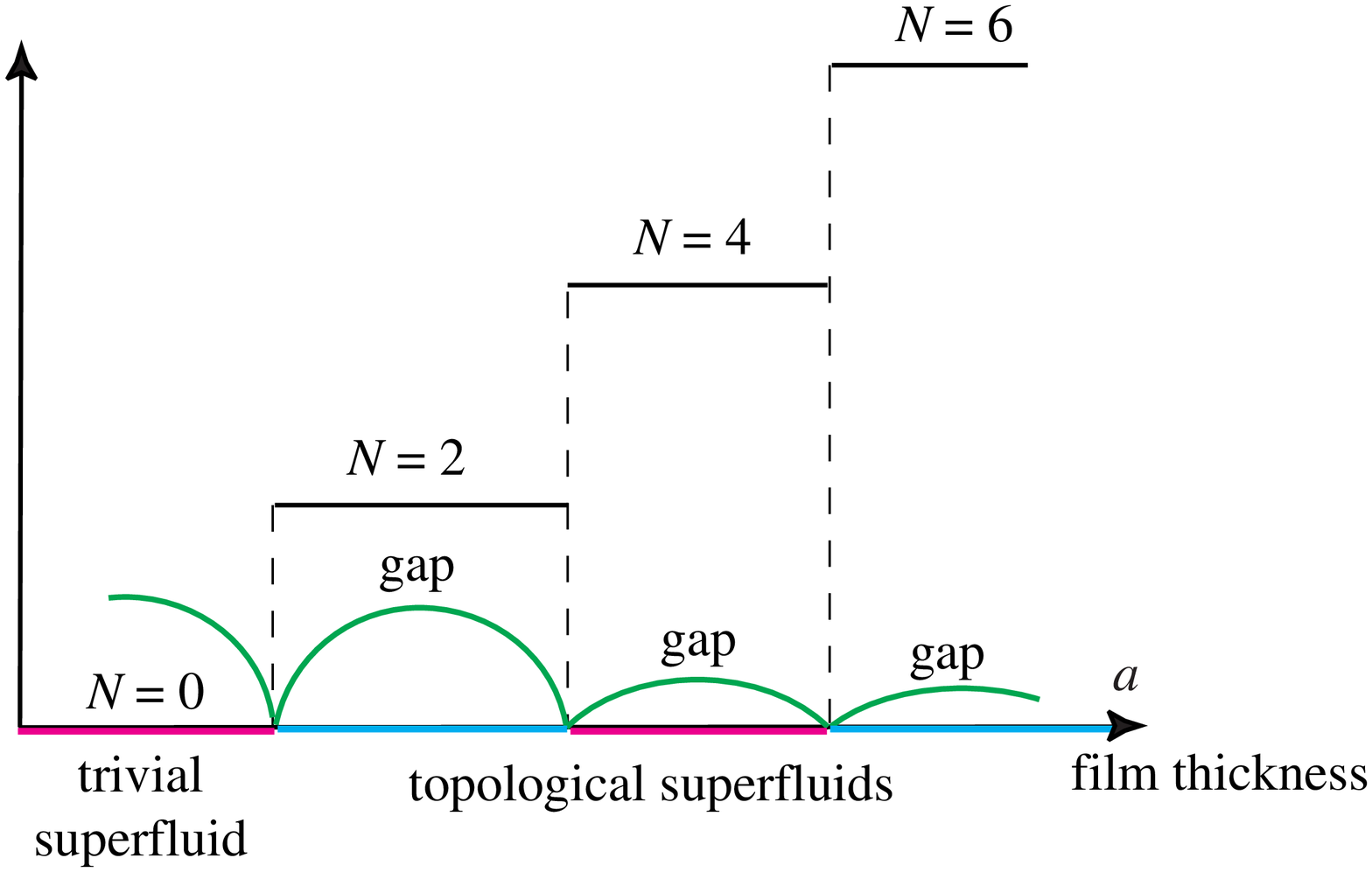}}
\caption{Dependence of the topological invariant (\ref{N2+1}) on the thickness of $^3$He-A film.
The even values of $N$ result from the spin degeneracy.
At the topological phase transitions between the states with different $N$, the gap in the spectrum of fermions is nullified. The interface between the vacua with different topological charges
contains gapless fermions -- edge states. The number of the gapless fermions is related to the difference
of the topological charges by index theorem.
}
\label{QPT_Fig}
\end{figure}

In general case of multilayered $^3$He-A,  topological charge $N$ may take any integer value of group $Z$. This charge determines  quantization of Hall and spin-Hall conductance 
and  the quantum statistics of the topological objects --  real-space skyrmions \cite{Volovik1988,VolovikYakovenko1989,SQHE,Volovik1992b}.
For $N=4k+1$ and   $N=4k+3$, skyrmion is anyon;  
for $N=4k+2$ it is fermion; and for $N=4k$ it is boson \cite{Volovik1992b}. This demonstrates the importance of the $Z_2$ and $Z_4$ subgroups of the group $Z$ in classification of topological matter;
and also  provides an example of the interplay of momentum-space and real-space topologies.  

\subsubsection{Planar phase: time reversal invariant gapped vacuum}

In case when some symmetry is present, additional invariants  appear, which correspond to dimensional reduction of topological invariant supported by symmetry $N_K$ in (\ref{MasslessTopInvariantStandard Model}):
\begin{equation}
 N_K= {e_{ijk}\over{24\pi^2}} ~
{\bf tr}\left[  \int    d^2p d\omega
~K G\partial_{p_i} G^{-1}
G\partial_{p_j} G^{-1}G\partial_{p_k}  G^{-1}\right],
\label{N2+1prime}
\end{equation}
where the matrix $K$ commutes with the Green's function matrix.
Example of the symmetric $2+1$ gapped state with  $N_K$ is the film of the planar phase of superfluid $^3$He \cite{VolovikYakovenko1989,Volovik1992b}. In the single layer case, the simplest expression for the Green's function is
\begin{equation}
  G^{-1}(\omega,p_x,p_y)=
 i \omega - H~~,~~ H= \tau_3\left(\frac{p_x^2+p_y^2}{2m} -\mu\right) + \tau_1 (\sigma_x p_x  + \sigma_yp_y)~,
 \label{planar_state}
\end{equation}
where $H$ is the Bogoliubov - de Gennes 
Hamiltonian for fermions in this spin-triplet $p$-wave superfluid. The symmetry operator $K$, which supports the topological invariant is $K=\tau_3\sigma_z$: it commutes with the Green's function. The
planar  state is time reversal invariant. 
It has trivial conventional topological charge $N=0$ and non-zero symmetry protected charge $N_K=2$. For the general case of the quasi 2D  film with multiple layers of the planar phase, the invariant $N_K$ belongs to the group $Z$.  The magnetic solid state analog of the planar phase is the 2D time reversal invariant topological insulator, which experiences the quantum spin Hall effect without external magnetic field \cite{HasanKane2010}.

 \section{Relativistic quantum vacuum and superfluid $^3$He-B}

Let us now turn to the class of 3+1 fully gapped systems, which is represented by Standard Model in its massive phase and superfluid $^3$He-B.

 In the broken symmetry phase of Standard Model below the electroweak transition, symmetry does not support the topological invariant responsible for the nodes in spectrum. In this phase there is no mass protection by topology and thus all the fermions become massive, i.e. Standard Model vacuum becomes the fully gapped insulator. 
 
 In quantum liquids, the fully gapped three-dimensional system, which is similar to the vacuum of massive Standard Model,  is represented by another phase of superfluid $^3$He -- the $^3$He-B. This phase has time reversal symmetry and nontrivial topology supported by symmetry, which gives rise  to the 2D gapless quasiparticles living at interfaces between vacua with different values of the topological invariant or on the surface of $^3$He-B 
 \cite{SalomaaVolovik1988,Volovik2009,Volovik2009b,Volovik2010}. 
 
 \subsection{Superfluid $^3$He-B }

 $^3$He-B belongs to the same topological class as the vacuum of Standard Model in its present insulating phase \cite{Volovik2010a}.  The topological classes of the $^3$He-B states can be represented by the following simplified Green's function and the Bogoliubov - de Gennes Hamiltonian:
 \begin{equation}
G^{-1}(\omega,{\bf p})=i\omega -H ~~,~~H=  \tau_3\left(\frac{p^2}{2m} - \mu\right)+  \tau_1
c^B{\mbox{\boldmath$\sigma$}}\cdot{\bf p}
\,.
\label{eq:B-phase}
\end{equation}
 In the limit $1/m=0$ this model $^3$He-B  transforms to the vacuum of massive relativistic Dirac particles with speed of light $c=c^B$ and mass parameter $M=-\mu$.

In the fully gapped  systems, the Green's function has no singularities in the  whole 4-dimensional space $(\omega,{\bf p})$. That is why we are able to use the Green's function at $\omega=0$,
which corresponds to the effective Hamiltonian, $H_{\rm eff}({\bf p})=-G^{-1}(0,{\bf p})$. The topological invariant relevant  for $^3$He-B and for quantum vacuum with massive Dirac fermions is:
\begin{equation}
N_K = {e_{ijk}\over{24\pi^2}} ~
{\bf tr}\left[  \int_{\omega=0}   d^3p ~K
~G\partial_{p_i} G^{-1}
G\partial_{p_j} G^{-1} G\partial_{p_k} G^{-1}\right]\,.
\label{3DTopInvariant_tau}
\end{equation} 
 with matrix $K=\tau_2$ which anti-commutes with the Green's function at $\omega=0$. In $^3$He-B, the $\tau_2$ symmetry   is combination of time reversal  and particle-hole symmetries; for Standard Model  the matrix  $\tau_2=\gamma_5\gamma^0$.  
Note that at $\omega=0$ the symmetry of the Green's function is enhanced, and thus there are more matrices $K$, which commute or anti-commute with the Green's function, than at $\omega\neq 0$.

 \begin{figure}
 \begin{center}
 \includegraphics[%
  width=0.6\linewidth,
  keepaspectratio]{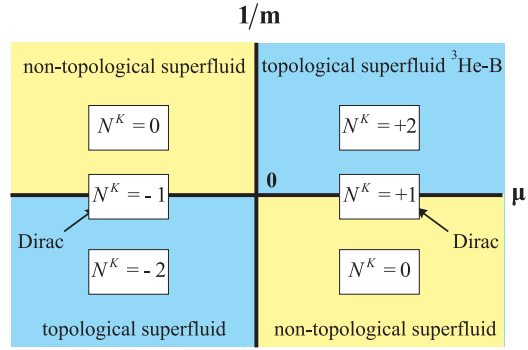}
\end{center}
  \caption{\label{3He-B}  Phase diagram of topological states of $^3$He-B in Eq.(\ref{eq:B-phase}) in the plane $(\mu,1/m)$. States on the line
  $1/m=0$ correspond to the  Dirac vacua, which Hamiltonian is non-compact. Topological charge of the Dirac fermions
  is intermediate between charges of compact $^3$He-B states.
The line $1/m=0$ separates the states with different asymptotic behavior of the Green's function at infinity:
$G^{-1}(\omega=0,{\bf p}) \rightarrow \pm \tau_3 p^2/2m$.
 The line $\mu=0$ marks topological quantum phase transition, which occurs between the weak coupling $^3$He-B
 (with $\mu>0$,
 $m>0$ and topological charge $N_K=2$) and the strong coupling $^3$He-B   (with $\mu<0$, $m>0$ and $N_K=0$).
  This transition is topologically equivalent to quantum phase transition between Dirac vacua with opposite mass parameter
 $M=\pm |\mu|$, which occurs when $\mu$ crosses zero along the line $1/m=0$.
 The interface which separates two states contains single Majorana fermion in case of $^3$He-B, and single chiral fermion
 in case of  relativistic quantum fields.  Difference in the nature of the fermions is that in Fermi superfluids and in superconductors
  the components of the Bogoliubov-Nambu spinor are related by complex conjugation. This reduces the number of degrees of freedom compared
  to Dirac case.
 }
\end{figure}

Fig. \ref{3He-B} shows the phase diagram of topological states of $^3$He-B in the plane $(\mu,1/m)$.
The line  $1/m=0$ corresponds to the Dirac vacuum of massive fermions, whose topological charge  is determined by the sign of mass parameter $M=-\mu$:
\begin{equation}
N_K= {\rm sign}(M)
\,.
\label{eq:DiracInvariants}
\end{equation}

The real superfluid $^3$He-B lives in the  corner of the phase diagram  $\mu>0$, $m>0$, $\mu\gg mc_B^2$, which also corresponds to the limit of the weakly interacting gas of  $^3$He atoms,
where the superfluid state is described by Bardeen-Cooper- Schrieffer (BCS) theory. However, in the ultracold Fermi gases with triplet pairing
 the strong coupling limit is possible near the Feshbach resonance \cite{GurarieRadzihovsky2007}.
 When $\mu$ crosses zero the topological quantum phase transition occurs, at which the topological charge $N_K$ changes from  $N_K=2$ to  $N_K=0$. The latter regime with trivial topology also includes
 the Bose-Eistein condensate (BEC) of two-atomic molecules. In other words, the BCS-BEC crossover in this system is always accompanied by the topological quantum phase transition, at which the topological invariant changes.
 
There is an important difference between $^3$He-B and Dirac vacuum. The space of the Green's function of free  Dirac fermions is non-compact: $G$ has different asymptotes at $|{\bf p}|\rightarrow \infty $ for different directions of momentum ${\bf p}$.   As a result, the topological charge of the interacting Dirac fermions depends on the regularization at large momentum. $^3$He-B can serve as regularization of the Dirac vacuum, which can be made in the Lorentz invariant way \cite{Volovik2010a}. One can see from Fig. \ref{3He-B}, that  the topological charge of free Dirac vacuum has intermediate value between the charges of the  $^3$He-B vacua with compact  Green's function. On the marginal behavior of free Dirac fermions see Refs. \cite{Haldane1988,Schnyder2008,Volovik2003,Volovik2009b}.

   The vertical axis separates the states with the same asymptote of the Green's function at infinity. The abrupt change of the topological charge across the line, $\Delta N_K=2$, with fixed asymptote shows that one cannot cross the transition  line adiabatically. This means that all the intermediate states on the line of this  QPT  are necessarily gapless. For the intermediate state between the free Dirac vacua with opposite mass parameter 
 $M$ this is well known. But this is applicable to the general case with or without relativistic invariance:
 the gaplessness is protected 
by the difference of topological invariants on two sides of transition.
 
 \subsection{From superfluid relativistic medium to $^3$He-B}

We have already seen, that the  Dirac
vacuum of massive Standard Model particles has nontrivial
topology. As a result the domain wall separating vacua with opposite signs
of the mass parameter $M$ contains fermion zero modes
\cite{JackiwRebbi1976}. 
Other examples of the topologically nontrivial states in relativistic theories
can be provided by dense quark matter, where chiral and color superconductivity is
possible. The topological properties of such fermionic systems
have been recently discussed in Ref.  \cite{Nishida2010}. In
particular,  in some range of parameters the isotropic
 triplet relativistic superconductor is topological and has the fermion zero modes both at the boundary and in
  the vortex core. On the other hand, there is a range of parameters, where this  triplet superconductor is
  reduced to the non-relativistic superfluid $^3$He-B \cite{Ohsaku2001}. That is why
the analysis in Ref.  \cite{Nishida2010} is applicable to $^3$He-B
and becomes particularly useful when the fermions living in the
vortex core are discussed.

 In relativistic superconductor or superfluid with the isotropic pairing -- such as  color superconductor in
  quark matter -- the fermionic spectrum is determined by Hamiltonian
  \begin{equation}
H=\tau_3\left( c {\mbox{\boldmath$\alpha$}}\cdot{\bf p} +  \beta M - \mu_R\right)+ \tau_1 \Delta
\,,
\label{eq:B-phaseRelatSinglet}
\end{equation}
for spin singlet pairing, and  by Hamiltonian
 \begin{equation}
H=\tau_3\left( c {\mbox{\boldmath$\alpha$}}\cdot{\bf p} +  \beta M - \mu_R\right)+ \gamma_5 \tau_1 \Delta
\,,
\label{eq:B-phaseRelat}
\end{equation}
for spin triplet pairing  \cite{Ohsaku2001,Nishida2010}.
Here $\alpha^i$, $\beta$ and $\gamma_5$ are Dirac matrices,
which in standard representation are
\begin{equation}
{\mbox{\boldmath$\alpha$}}  = \left( \begin{array}{cc}
0 & {\mbox{\boldmath$\sigma$}} \\
  {\mbox{\boldmath$\sigma$}} & 0 \end{array} \right)
  \;\;\;\; ,    \;\;\;\; \beta=\left( \begin{array}{cc}1 & 0 \\
   0 & -1 \end{array} \right)
    \;\;\;\; ,    \;\;\;\; \gamma_5=\left( \begin{array}{cc}0 & 1 \\
   1 & 0  \end{array} \right);
  \end{equation}
  $M$ is the rest energy of fermions; $\mu_R$ is their relativistic chemical potential as distinct
  from the non-relativistic chemical potential $\mu$;
  $\tau_a$ are matrices in Bogoliubov-Nambu space;
and $\Delta$ is the gap parameter.

 In non-relativistic limit the low-energy Hamiltonian is obtained by standard procedure, see e.g. \cite{Nishida2010b}. The non-relativistic limit is determined by the conditions
\begin{equation}\label{cond1}
 cp\ll M
\end{equation}
and
 \begin{equation}\label{cond2}
 |M-\sqrt{\mu_R^2+\Delta^2}|\ll M\,.
 \end{equation}
Under these conditions the Hamiltonian (\ref{eq:B-phaseRelatSinglet}) reduces to the
Bogoliubov - de Gennes (BdG)
Hamiltonian for fermions in spin-singlet $s$-wave superconductors, while (\ref{eq:B-phaseRelat})  transforms to the
BdG Hamiltonian relevant for fermions in isotropic spin-triplet $p$-wave superfluid $^3$He-B
in Eq.(\ref{eq:B-phase}):
\begin{equation}
H=\tau_3\left(\frac{p^2}{2m} - \mu\right)+  c^B\tau_1{\mbox{\boldmath$\sigma$}} \cdot{\bf p}  ~~,~~ m=\frac{M}{c^2}~~,~~c^B=c\frac{\Delta}{M}
\,,
\label{eq:B-phase2}
\end{equation}
 where the nonrelativistic chemical potential $\mu=\sqrt{\mu_R^2+\Delta^2}- M$.

 \begin{figure}
 \begin{center}
 \includegraphics[%
  width=0.8\linewidth,
  keepaspectratio]{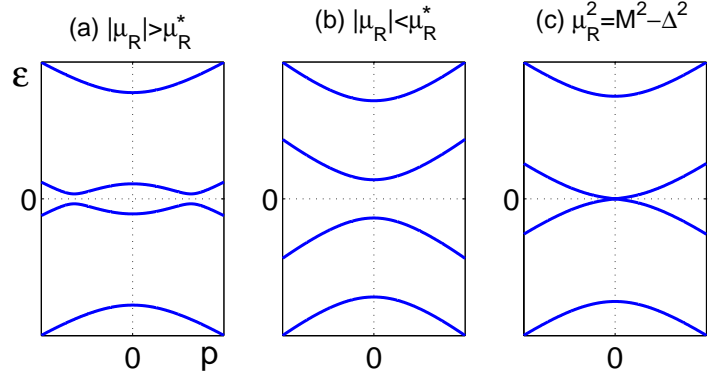}
\end{center}
 \caption{\label{PlotSpectrum} Plot of the spectrum of relativistic Hamiltonian
  (\ref{eq:B-phaseRelat})
 for two generic cases: (a) $|\mu_R| > \mu_R^*$ when the minimum in the energy spectrum is away from the origin and (b) 
 $|\mu_R| < \mu_R^*$  when the minimum in the energy spectrum is at ${\bf p}=0$. At $|\mu_R| > \mu_R^*$ there is a soft quantum phase transition between these two vacua. This transition is not topological, and thus the gap in the energy spectrum does not close at the transition. The gap  closes at 
 the topological transition occurring at $\mu_R^2=M^2-\Delta^2$ as shown in plot (c).}
\end{figure}

 The  Dirac-BdG system in Eq.(\ref{eq:B-phaseRelat}) has the
 following spectrum
 \begin{equation}\label{SpectrumGen}
  \varepsilon=\pm\sqrt{M^2+c^2p^2+\Delta^2+\mu_R^2\pm 2
  \sqrt{M^2(\mu_R^2+\Delta^2)+\mu_R^2c^2p^2}}.
 \end{equation}
 This spectrum is plotted in
Fig.\ref{PlotSpectrum}.
 Depending on the value of the parameters $\mu_R$, $\Delta$, $M$
 the spectral branches have different
 configurations.

There is a soft quantum phase transition, at which the
position of the minimum of energy $E(p)$ shifts from the origin ${\bf p}=0$,
and the energy profile forms the Mexican hat in momentum space.
This momentum-space analog of the Higgs transition  \cite{Volovik2007}
occurs when the relativistic chemical potential
  $\mu_R$ exceeds the critical value
  \begin{equation}
\mu^*_R=\left(\frac{M^2}{2} + \sqrt{ \frac{M^4}{4}
+M^2 \Delta^2}\right)^{1/2} \,. \label{eq:SoftTransition}
\end{equation}
 Figures \ref{PlotSpectrum} (a) and  \ref{PlotSpectrum} (b) demonstrate two generic cases:
 $|\mu_R| > \mu_R^*$
 when there are extremums of function $\varepsilon (p)$ at $p\neq 0$ and
  $|\mu_R| < \mu_R^*$ when all
  extremums are at the point $p=0$.
The formation of the  Mexican hat at $|\mu_R| = \mu_R^*$ is an example of non-topological
 quantum phase transition, which occurs without change of the topological invariants and thus is not accompanied by the gapless intermediate state.
Let us turn to the topological quantum phase transitions, at which the ${\bf p}$-space topological invariant changes and the gap closes at the transition point as is shown in Fig.  \ref{PlotSpectrum} (c).

\subsection{Topology of relativistic medium and  $^3$He-B}

\begin{figure}
 \begin{center}
 \includegraphics[%
  width=0.6\linewidth,
  keepaspectratio]{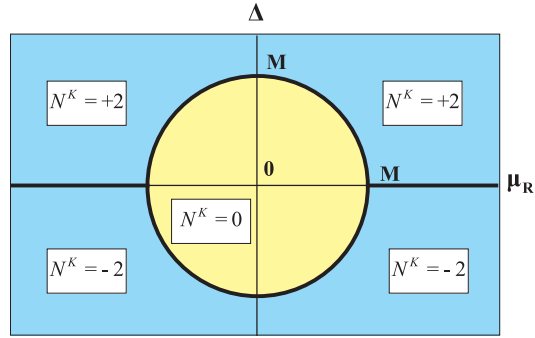}
\end{center}
  \caption{\label{PhaseDiagramRel}  Phase diagram of ground states of relativistic
 triplet superfluid in Eq.(\ref{eq:B-phaseRelat}) in the plane $(\mu_R, \Delta)$. Topological quantum phase transitions are marked by
  thick lines. The states inside the circle $\mu_R^2+\Delta^2=M^2$ are topologically trivial.
 The states outside this circle represent topological superconductors. The states on the lines of
  topological quantum phase transition are gapless.
 }
\end{figure}

Fig. \ref{PhaseDiagramRel}  shows the phase diagram of the vacuum states of relativistic  triplet superconductors. Different vacuum states are characterized by different values of the topological invariant $N_K$ in Eq.(\ref{3DTopInvariant_tau}), where 
the Green's function matrix at zero
frequency $G^{-1}(\omega=0,{\bf p})$ is equivalent to effective  Hamiltonian. For $^3$He-B in Eq.(\ref{eq:B-phase2}) and for triplet relativistic superconductor in Eq. (\ref{eq:B-phaseRelat})
the relevant matrix $K=\tau_2$, which anti-commutes with the Hamiltonian.
The vacuum states
with different $N_K$ cannot be adiabatically connected, and thus at the phase transition lines the states are gapless. The circle $\mu_R^2+\Delta^2=M^2$ is an example of the line of  topological quantum phase transition. In non-relativistic limit this corresponds to the line $\mu=0$  in  Fig.\ref{3He-B}.  The states inside the circle $\mu_R^2+\Delta^2=M^2$ are topologically trivial, while
the states outside this circle represent topological superconductivity \cite{Nishida2010}. The vacuum states
  with $\mu_R^2+\Delta^2>M^2 $ and $\mu_R^2+\Delta^2<M^2$ cannot be adiabatically connected which leads to the gap closing in   Fig.\ref{PlotSpectrum} (c). Discontinuity in the topological charge across the transition induces discontinuity in the energy of the ground state across the transition. For example, for the 2+1 $p_x+ip_y$ superfluid/superconductor the quantum phase transition is of third order, meaning that the third-order derivative of the ground state
energy is discontinuous \cite{Rombouts2010}.

 \section{Fermions in the core of strings in topological materials}

In relativistic theories there is an index theorem which relates the number of fermion zero modes localized on a vortex with the vortex winding number \cite{JackiwRossi1981}. We know that the Dirac vacuum considered in Ref.  \cite{JackiwRossi1981} has nonzero  topological charge. This suggests that the existence of zero energy states in the core is sensitive not only to the real-space topological charge of a vortex, but also to the momentum-space topological charge of the quantum vacuum in which the vortex exists, and if so the index theorem should be extended to vortices in any fully gapped systems, including the non-relativistic superfluid $^3$He-B. Here we discuss this issue of the connection between the topological charge of the vacuum and existence of the fermion zero modes on the topological objects in this vacuum. Our examples demonstrate that during the topological quantum phase transition
from the topologically nontrivial vacuum to the trivial one, the fermion zero modes on vortices (or other objects) disappear.

\subsection{Vortices in $^3$He-B and  relativistic strings}

Since in this problem both the momentum-space topology of bulk state and the real-space  topology of the vortex or other topological defects are involved, the combined topology of the Green's function in the coordinate-momentum space $(\omega,{\bf p},{\bf r})$  \cite{GrinevichVolovik1988,Volovik1991,Volovik2003,TeoKane2010a,TeoKane2010b} seems to be relevant.  However, though the bulk-vortex correspondence does evidently exist, the explicit index theorem which relates the existence of the fermion zero modes to the topological charge of the bulk state and the vortex winding number is still missing. 
The existing index theorems are applicable only to particular cases, see e.g.
\cite{Volovik1991,Nishida2010,TeoKane2010b,Herbut2010}. There is also a special index theorem for superconductors/superfluids  with a small gap $\Delta \ll \mu$. Spectrum of fermions in these superconductors has branches which cross zero energy as a function discrete quantum number -- angular momentum $L$ \cite{Caroli1964}. The index theorem relates the number of such branches with the vortex winding number \cite{Volovik1993a}. Here we are interested in the true fermion zero modes -- the branches of spectrum $E(p_z)$, which cross zero as function of momentum $p_z$ along the vortex line. 
 
 \begin{figure}
 \begin{center}
 \includegraphics[%
  width=0.8\linewidth,
  keepaspectratio]{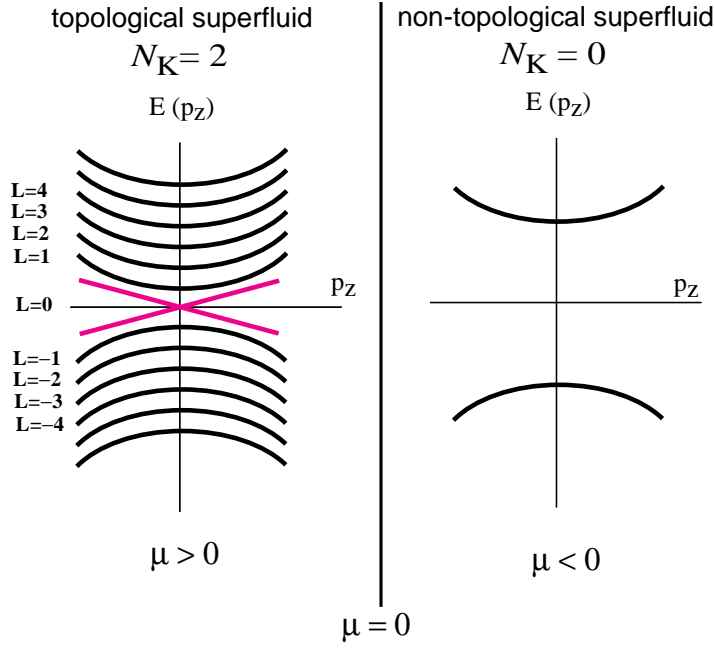}
\end{center}
  \caption{\label{FermionsBphaseVortex}
Schematic illustration of spectrum of  the fermionic bound states in the core
 of the most symmetric vortex with $n=1$, the so-called $o$-vortex \cite{SalomaaVolovik1987}, in fully gapped spin triplet superfluid/superconductor of $^3$He-B type.
({\it left}): Spectrum of bound state in the $^3$He-B, vortex which corresponds to the weak
coupling BCS regime with non-zero topological charge $N_K=2$
\cite{MisirpashaevVolovik1995}.   $L$ is the azimuthal quantum number of fermions in the vortex core. There are two fermion zero modes, which cross zero energy in the opposite directions.  ({\it right}): The same vortex but in the
topologically trivial BEC regime, $N_K=0$, does not have fermion zero modes. The spectrum
of bound states is fully gapped.  Fermion zero modes disappear at the topological quantum phase
transition, which occurs in bulk liquid at $\mu=0$. Similar situation may take place for strings
 in color superconductors in quark matter \cite{Nishida2010}.
 }
\end{figure}
 
Example, which demonstrates that the connection between the topological charge $N_K$ and the existence of Majorana fermions -- fermion zero modes on vortices -- is in Fig. \ref{FermionsBphaseVortex}.
 For $^3$He-B, which lives in the BCS  range of parameters where $N_K\neq 0$, the gapless fermions in the core have been
  found in Ref. \cite{MisirpashaevVolovik1995}. On the other hand, in the strong coupling limit
the   $^3$He-B transforms to the Bose-Einistein condensate (BEC) of molecules. The latter does not contain
fermionic excitations, and thus one should not expect the existence of gapless fermions in the 
vortex core. Thus one expects that somewhere in the region of crossover between the BCS-like
and BEC-like regime the spectrum of fermions localized on vortices must be reconstructed.
On the other hand, the topological reconstruction of the fermionic
  spectrum  in the vortex core cannot occur during the adiabatic deformation. The discontinuous deformation of the spectrum in the core is only possible during the topological quantum phase
  transition in bulk: at such transition the intermediate bulk gapless state is crossed which destroys the adiabaticity.
This is just what happens:   in the  BEC limit, the chemical potential $\mu$ is negative and the topological charge of the vacuum state is trivial,  $N_K=0$.  The reconstruction of the bulk spectrum at the topological quantum phase transition occuring at $\mu=0$, see Fig.
   \ref{3He-B}, triggers reconstruction of the spectrum of fermion zero modes in the core.
At $\mu<0$  the topological charge $N_K$ nullifies and simultaneously the gap in the spectrum of core fermions arises,
 see Fig. \ref{FermionsBphaseVortex}. The similar situation, when the quantum phase transition in bulk produces leads to appearance or disappearance of fermion zero modes, has been discussed in Ref. \cite{MizushimaMachida2010}
 for the other type of $p$-wave vortices; in Ref.  \cite{Nishida2010} for strings
 in color superconductors in quark matter; and in Refs.  \cite{Lutchyn2010,Oreg2010} for Majorana fermions on the edges of quantum wire (the review on Majorana fermions in superconductors can be found in Ref. \cite{Beenakker2011}). 

Another example is provided by the  fermions on relativistic vortices  in Dirac vacuum discussed
in Ref. \cite{JackiwRossi1981}. The Dirac vacuum has the nonzero topological invariant, $N_K=\pm 1$, see Fig. \ref{3He-B}.
This is consistent with the existence of the fermion zero modes on vortices, found in Ref. \cite{JackiwRossi1981}.
The index theorem for fermion zero modes on these vortices can be derived using the topology in combined coordinate
and momentum space. The number of fermion zero modes on a vortex $N_{\rm zm}$ can be expressed  via the 5-form topological invariant in terms of Green's function $G(\omega, {\bf p},{\bf r})$ \cite{SilaevVolovik2010,Shiozaki2011}
 \begin{equation}
N_{\rm zm}= \frac{1} {4\pi^3 i} ~
{\bf tr}\left[  \int   d^3p d\omega \oint_C dl \,
 G\partial_{p_x}G^{-1}
 G\partial_{p_y} G^{-1}
 G\partial_{p_z}G^{-1}
 G\partial_{\omega} G^{-1}
 G\partial_{l} G^{-1}\right] \,.
 \label{N5}
 \end{equation}
  The space integral is along the closed contour $C$ around the vortex line.
For the vortex in Dirac vacuum, equation (\ref{N5})  
 reproduces the index theorem discussed in Ref. \cite{JackiwRossi1981}: the algebraic number of fermion zero
 modes equals the vortex winding number $N_{\rm zm}=n$.

For vortex in $^3$He-B one obtains $N_{\rm zm}=0$. This is, however,  consistent with Fig. \ref{FermionsBphaseVortex}: two branches of zero modes have opposite signs of velocity $v_z=dE/dp_z$. Though the algebraic sum of zero modes,  $N_{\rm zm}=1-1=0$, due to the special symmetry of the vortex configuration the two branches of fermion zero modes do not cancel each and the gap is not formed.  To resolve the fermion zero modes in the systems
with symmetry, the index theorem for the zero modes must be supplemented  by symmetry consideration.
  
  The 5-form topological invariant similar to Eq.(\ref{N5})  has been discussed also in \cite{Volovik2003,ZhongWang2010}. In particular it is responsible for the topological stability of the 3+1 chiral fermions emerging  in the core of the domain wall separating  topologically different  vacua in 4+1 systems (see Sec. 22.2.4 in \cite{Volovik2003}). The topological invariant for the general $2n+1$ insulating relativistic vacua and the bound chiral fermion zero modes emerging there have been considered in \cite{Kaplan1992,Golterman1993,Kaplan2011}. 
  Application of the 5-form topological invariant to the states in lattice chromodynamics can be found
  in \cite{ZubkovVolovik2012,Zubkov2012b}.

 \subsection{Flat band in a vortex core: analog of Dirac string terminating on monopole}

\begin{figure*}
\includegraphics[width=0.5\textwidth]{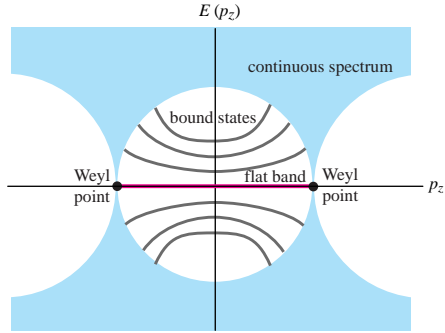}
\caption{Schematic illustration of the spectrum of bound states $E(p_z)$ in the vortex core. The branches of bound states terminate at points where their spectrum merges with the continuous spectrum in the bulk.
The flat band terminates at points where the spectrum has zeroes in the bulk, i.e. when it merges with Weyl points. It is the ${\bf p}$-space analog of a Dirac string terminating on a monopole, another analog is given by  the Fermi arc in Fig. 1 {\it bottom right} .} 
\label{VortexFlatBand}
\end{figure*}

The topological bulk-vortex correspondence exists also for vortices in gapless vacua. The
topological protection of fermion zero modes is provided by the nontrivial topology of
three-dimensional Weyl points in the bulk. This bulk-vortex
correspondence \cite{Volovik2010} is illustrated in Fig.
\ref{VortexFlatBand}. In bulk there is a pair
of Weyl points with opposite topological charges $N=\pm 1$ in Eq.(\ref{MasslessTopInvariant3D}). The
projections of these Weyl points on the direction of the vortex
line determine the boundaries of the region where the spectrum of
fermions bound to the vortex core is exactly zero,
$E(p_z)=0$, for all $p_z$ within this region. Such flat band was first obtained
in Ref.~\cite{KopninSalomaa1991} for the noninteracting model. However, due to the topological protection, it is not destroyed by interactions. The spectrum of bound states 
in a singly quantized vortex in $^3$He-A is shown in Fig. \ref{VortexFlatBand}.
The 1D flat band terminates at points where the spectrum of bound state merges with zeroes in the bulk, i.e. with Weyl points.

\section{Discussion}

 The last decades demonstrated that topology becomes a very important tool in physics. Topology in momentum space is the main characteristics of  the ground states of a system at zero temperature ($T=0$), in other words it is the characteristics of quantum vacua. The gaplessness of fermions in bulk, on the surface or inside the vortex core is protected by topology, and thus is not sensitive to the details of the microscopic physics (atomic or trans-Planckian). Irrespective of the deformation of the parameters of the microscopic theory, the value of the gap (mass) in the energy spectrum of these fermions remains strictly zero. This solves the main hierarchy problem in particle physics: for  fermionic vacua with Fermi points the masses of elementary particles are naturally small.

 The vacua, which have nontrivial topology in momentum space, are called the topological matter, and the quantum vacuum of Standard Model is the representative of the topological matter alongside with topological superfluids and superconductors, topological insulators and semi-metals, etc.
 There is a number of of topological invariants in momentum space of different dimensions. They determine universality classes of the topological matter and the type of the effective theory which  emerges at low energy and low temperature. In many cases they also give rise to emergent symmetries, including the effective Lorentz invariance and probably all the symmetries of Standard Model, and  emergent phenomena such as gauge and gravitational fields. The symmetry appears to be the  secondary factor, which emerges in the low-energy corner due to topology, and it is possible that it is topology of the quantum vacuum, which is responsible for the properties of the  fermionic matter in the present low-energy Universe.
 
 The topological invariants in extended momentum and coordinate space determine the bulk-surface and bulk-vortex correspondence. They connect the momentum space topology in bulk with
 the real space. These invariants determine the fermion zero modes living on the surface of a system or in the core of topological defects (vortices, strings, domain walls, solitons, hedgehogs, etc.).

In respect to gravity, the momentum space topology gives some lessons. First, in the effective gravity emerging at low energy, the collective variables represent the tetrad field and spin connections. In this approach the metric field emerges as the composite object of tetrad field, and thus the Einstein-Cartan-Sciama-Kibble theory with torsion field is more relevant for the description of gravity
(see also \cite{Akama1978,Volovik1986,Wetterich2004,Diakonov2011}).

Second, the topology suggests several scenarios of Lorentz invariance violation governed by topology.
Among them the splitting of Fermi point and development of the Dirac points with quadratic and cubic spectrum. The latter leads to the natural emergence of the Ho\v{r}ava-Lifshitz gravity.

\begin{acknowledgement}
 This work
is supported in part by the Academy of Finland and its COE program
2006--2011
\end{acknowledgement}

\end{document}